\documentclass[acmsmall]{acmart}

\AtBeginDocument{%
  \providecommand\BibTeX{{%
    \normalfont B\kern-0.5em{\scshape i\kern-0.25em b}\kern-0.8em\TeX}}}

\setcopyright{acmcopyright}
\copyrightyear{2023}
\acmYear{2023}
\acmDOI{XXXXXXX.XXXXXXX}

 \acmJournal{JACM}
 \acmVolume{37}
 \acmNumber{4}
 \acmMonth{5}

\usepackage{booktabs}
\usepackage{graphicx}
\usepackage{fullpage}
   \usepackage{color}
\usepackage {natbib}
\usepackage{fancyhdr, lipsum}
\usepackage{amsmath} 
 \usepackage{pgf-pie}
 \usepackage{pgfplots} 
\pgfplotsset{compat=1.8} 
\usepackage{graphicx} 
\usepackage{xcolor} 
\usepackage{tabularx}
\usepackage{longtable}
\usepackage{array}
\usepackage{fontawesome}
\usepackage{tikz}
\usepackage [T1] {fontenc}
\usepackage{multirow}
\usepackage{appendix}
\usepackage{float}
\def\mybarhhigh#1#2{
   {\color{black}\rule{#1mm}{4pt}}  #2}

\begin{document}

\title{Ethics in the Age of AI: An Analysis of AI Practitioners' Awareness and Challenges}

\author{Aastha Pant}
\email{aastha.pant@monash.edu}
\authornotemark[1]
\affiliation{%
  \institution{Monash University}
  \streetaddress{Wellington Road}
  \city{Melbourne}
  \state{Victoria}
  \country{Australia}
  \postcode{3800}
}

\author{Rashina Hoda}
\affiliation{%
  \institution{Monash University}
  \streetaddress{Wellington Road}
  \city{Melbourne}
  \country{Australia}}
\email{rashina.hoda@monash.edu}

\author{Simone V. Spiegler}
\affiliation{%
  \institution{Monash University}
  \streetaddress{Wellington Road}
  \city{Melbourne}
  \state{Victoria}
  \country{Australia}
  \postcode{3800}}
\email{simone.spiegler@monash.edu}

\author{Chakkrit Tantithamthavorn}
\affiliation{%
  \institution{Monash University}
  \city{Melbourne}
  \country{Australia}}
  \email{chakkrit@monash.edu}

\author{Burak Turhan}
\affiliation{%
  \institution{University of Oulu}
  \city{Oulu}
  \country{Finland}}
  \email{burak.turhan@oulu.fi}

\renewcommand{\shortauthors}{Pant, et al.}

\begin{abstract}
Ethics in AI has become a debated topic of public and expert discourse in recent years. But what do people who build AI -- AI practitioners -- have to say about their understanding of AI ethics and the challenges associated with incorporating it in the AI-based systems they develop? Understanding AI practitioners' views on AI ethics is important as they are the ones closest to the AI systems and can bring about changes and improvements. We conducted a survey aimed at understanding AI practitioners' \emph{awareness} of AI ethics and their \emph{challenges} in incorporating ethics. Based on 100 AI practitioners' responses, our findings indicate that majority of AI practitioners had a \emph{reasonable} familiarity with the concept of AI ethics, primarily due to \emph{workplace rules and policies}. \emph{Privacy protection and security} was the ethical principle that majority of them were aware of. Formal education/training was considered \emph{somewhat} helpful in preparing practitioners to incorporate AI ethics. 
The challenges that AI practitioners faced in the development of \textit{ethical} AI-based systems included (i) general challenges, (ii) technology-related challenges and (iii) human-related challenges. We also identified areas needing further investigation and provided recommendations to assist AI practitioners and companies in incorporating ethics into AI development.
 
\end{abstract}



\maketitle

\section{Introduction}
AI technology is becoming increasingly pervasive across industries and contexts, making AI ethics more important than ever \citep{mittelstadt2019principles, hagendorff2020ethics}. The term `AI ethics' is defined as, \emph{``the principles of developing AI to interact with other AIs and humans ethically and function ethically in society"} \citep{siau2020artificial}. The importance of ethical consideration in AI has been highlighted by several incidents in recent years \citep{bostrom2018ethics}. For example, there have been ethical issues such as Google's Machine learning (ML) algorithm that was gender-biased against women as it associated men with Science, Technology, Engineering, and Mathematics (STEM) careers more frequently than women \citep{prates2020assessing}; GitHub’s use of copyrighted source code as training data for their ML-powered GitHub Copilot which is an example of a privacy issue \citep{twitter}; Amazon's AI-powered recruitment tool that was gender-biased \citep{Amazon}. All these incidents demonstrate the repercussions of neglecting ethics in AI development, emphasising the necessity of incorporating ethical considerations in AI to prevent similar incidents in the future.

The extensive use of AI-based systems in various fields and the ethical incidents of AI-based systems have increased the number of studies in this area. Various studies have been conducted in the area of AI ethics and most studies are theoretical and conceptual in nature \citep{chivukula2021identity}. For instance, studies have been carried out to analyse ethical principles of AI \citep{mark2019ethics, siau2020artificial, jobin2019global, fjeld2020principled}. Studies have been conducted with the aim of enhancing the development of ethical AI-based systems or minimising ethical issues that are highly prevalent in the current era. Studies have also proposed several frameworks, models, and methods to assist AI practitioners to incorporate ethics into AI-based systems such as maturity model \citep{vakkuri2021time}, ECCOLA method \citep{vakkuri2020eccola}, AI ethics framework \citep{etzioni2017incorporating}. Likewise, several AI ethical guidelines and principles have been developed and formulated \citep{Google, IBM, Microsoft}. Despite the abundance of guidelines and studies in the field of AI ethics, various ethical problems associated with AI systems continue to persist. The mere development of guidelines is insufficient; AI practitioners\footnote{The term `practitioners' in our study includes AI developers, engineers, specialists, experts, designers, and stakeholders involved in the design and development activities of AI-based systems. The terms `AI practitioners' and `AI developers' are used interchangeably throughout our study.} must also adhere to them during the development of AI-based systems. It is therefore imperative to examine whether they are following these guidelines and taking the necessary steps in the development of such systems. While conducting theoretical research is crucial, it's equally essential to conduct empirical studies to understand the views of AI practitioners on AI ethics since it ultimately relies on the practitioners to adhere to ethical principles of AI \citep{vakkuri2020current}. 

Recent studies have discussed several aspects of AI practitioners' awareness of ethics in AI. For example, \citet{vakkuri2019implementing, stahl2022organisational} discussed that AI practitioners were aware of the concept of ethics in AI, its importance, and its relevance. Several tools and methods have been developed to raise the awareness of AI ethics among AI practitioners \citep{morley2021ethics}. Likewise, studies have been conducted to focus on investigating the challenges related to adhering to specific ethical principles of AI such as \emph{fairness}, \emph{accountability}, and \emph{privacy} \citep{orr2020attributions, madaio2022assessing, rakova2021responsible}. The significance of AI ethical principles and challenges associated with the implementation of those principles have been conducted through an empirical investigation with AI practitioners and lawmakers \citep{khan2022ai}. Although several studies have reported the challenges of AI practitioners in incorporating ethics in AI, most of those studies focused on investigating the challenges of AI practitioners and other stakeholders related to specific ethical principles of AI. There is a lack of research that focuses on investigating the overall challenges of AI practitioners in incorporating ethics in AI.

Previously, we conducted a Grounded Theory Literature Review (GTLR) of 38 empirical articles to gain insights into research studies that focused on investigating AI practitioners' views on AI ethics and developed a \emph{taxonomy of ethics in AI from developers' viewpoints} spanning five categories: (i) \emph{developer awareness}, (ii) \emph{developer perception}, (iii) \emph{developer need} (iv) \emph{developer challenge}, and (v) \emph{developer approach} \citep{pant2022AIethics}. In this study, we conducted a \textbf{survey} with AI practitioners to investigate two aspects (categories) including, (i) \emph{developer awareness} of ethics in AI and (ii) \emph{developer challenge} in incorporating ethics in AI that we derived through our GTLR. The two research questions (RQs) of this study are:

\textbf{RQ1. How aware are AI practitioners of different aspects related to AI ethics?} \\
To answer this RQ, we investigated: (i) the extent to which AI practitioners are aware of the `AI ethics’ concept, (ii) what ethical principles AI practitioners are aware of, (iii) reasons for AI practitioners' awareness of ethics and (iv) awareness of the role of formal education/training in preparing AI practitioners to incorporate ethics in AI.

\textbf{RQ2. What challenges/barriers do AI practitioners face in incorporating ethics in AI?} \\
To answer this RQ, we investigated the degree of challenges AI practitioners face in considering and following each AI ethical principle using Australia's AI Ethics Principles\footnotemark{} as a reference. We also explored the overall key challenges/barriers that AI practitioners face in incorporating ethics in AI.
\footnotetext{https://www.industry.gov.au/publications/australias-artificial-intelligence-ethics-framework/australias-ai-ethics-principles}
Our survey contained 15 questions in total (12 closed-ended and 3 open-ended questions). We collected data in two rounds. In the first round, we advertised our survey on social media platforms such as LinkedIn and Twitter and obtained 17 responses, and in the second round, we collected the data for our study using the Prolific platform and obtained 83 responses, making the total number of responses to our study 100. We used \emph{Socio-Technical Grounded Theory (STGT) method for data analysis}~\cite{hoda2021socio} to analyse the qualitative data and descriptive statistics to analyse the quantitative data. The main contribution of our study is that we designed a set of recommendations for further research on AI practitioners' \emph{awareness} and \emph{challenges} around ethics in AI that would be beneficial for both AI practitioners and AI researchers to enhance AI practitioners' awareness of AI ethics and the challenges in incorporating ethics in AI. We also provide recommendations to AI educators.

The rest of the paper is organised as follows. Section \ref{Sec:RW} presents the related work followed by the research methodology in Section \ref{Sec:Methodology}. We present the findings in Section \ref{Findings} followed by a discussion on key findings and recommendations in Section \ref{sec:Discussion}. Then, we provide the limitations and threats to the validity of our study in Section \ref{sec:Limitations} which is followed by a conclusion in Section \ref{sec:conclusion}. 

\section{Related Work} \label{Sec:RW}
In this section, we provide a summary of the previous research.

\subsection{AI Practitioners' Awareness of Ethics in AI}
Studies have been conducted to investigate the awareness of AI practitioners of ethics and ethical principles of AI. The majority of these studies have focused on understanding if AI practitioners are aware of the concept of AI ethics and its importance in AI development. \citet{vakkuri2019ethically} conducted semi-structured interviews with AI practitioners and concluded that they were aware of the concept of \emph{ethics} in AI and its importance. \citet{stahl2022organisational} applied a case study research strategy and reported that AI practitioners were aware of the relevance of \emph{ethics} in AI. \citet{govia2020coproduction} conducted interviews and field observations with AI specialists to investigate the social or ethical implications of AI based on their working experience in AI development. They found that AI specialists were aware of the philosophical theories of AI ethics.  

Studies have also reported the awareness of AI practitioners on different ethical principles of AI. \citet{vakkuri2019ethically} conducted interviews with AI practitioners from five case companies to understand the practices used by them to incorporate ethics in AI. They found that the participants of all five case companies were aware of and concerned about the issues related to system \emph{transparency}, which is an AI ethical principle. \citet{christodoulou2021democracy} conducted six focus groups with 63 AI experts to investigate the ethical issues of digital media with a  focus on AI and Big Data. They reported that the participants were aware of \emph{`Transparency'} as an ethical principle of AI. \citet{vakkuri2019ethi} reported that AI practitioners had knowledge about the term `transparent AI' as it was one of their goals of AI development. \citet{mark2019ethics} discussed how participants were knowledgeable about the transparency law, which assisted them in identifying which data should be disclosed and which data should be kept confidential when creating an AI system and strived to create transparent systems. Likewise, \citet{holstein2019improving} conducted semi-structured interviews with 35 ML practitioners and a survey with 267 ML practitioners to investigate their challenges and needs to develop fair ML systems. They found that the majority of them were aware of \emph{`Fairness'} principle and possessed knowledge of fairness-related issues of ML systems. \citet{veale2018fairness} carried out interviews with 27 ML practitioners across 5 OECD countries regarding the challenges and design needs for algorithmic support in high-stakes public sector decision-making and found that the practitioners were aware of \emph{`Fairness'} AI ethical principle and worked towards abolishing fairness issues in AI systems. The participants were also aware of \emph{`Accountability/ Responsibility'} and its importance in the development and deployment of AI-based systems and took responsibility for any harm caused by their creations. 
\citet{ibanez2022operationalising} conducted 22 semi-structured interviews and 2 focus groups with AI practitioners to investigate how AI companies bridge the gap between AI ethical principles and practice. They found that \emph{`Privacy'} is the principle that AI practitioners were most aware of and extensively discussed, with data and information privacy being a significant concern for organisations. \citet{ryan2021research} conducted a multiple case study analysis to investigate the ethical concerns arising from the implementation and use of Big data and AI. They found that \emph{`Privacy'} was the only ethical principle that the participants of all 10 case studies were aware of and discussed ethical issues related to it. 

Most of the previous work has focused on understanding if AI practitioners are aware of the concept of AI ethics and understanding AI practitioners' awareness of specific ethical principles of AI. There still remain areas that need to be investigated when it comes to the awareness of AI practitioners on AI ethics so that effective strategies could be developed to enhance the awareness of AI ethics among AI practitioners. 

\subsection{AI Practitioners' Challenges in Incorporating Ethics in AI}

Studies have reported several challenges that AI practitioners face in incorporating ethics during the development of AI-based systems.  However, the main focus of these studies was not to investigate the specific challenges faced by AI practitioners in incorporating ethics in AI. For example, \citet{vakkuri2019ethically} and \citet{govia2020coproduction} conducted empirical studies to explore the practices used by AI practitioners in incorporating ethics into AI development and investigate the social and ethical implications of AI respectively. In doing that, they discovered several challenges discussed by the participants related to ethics incorporation in AI. \citet{orr2020attributions} conducted interviews with 21 AI practitioners to investigate how they attribute and distribute ethical responsibility for AI systems. During that process, several AI practitioners discussed the challenges they face in developing ethical AI systems including challenges related to organisational norms, legislative regulations, users, and AI machines. \citet{ibanez2022operationalising} conducted an empirical study to understand the gap between principles and practices in AI ethics. They also explored various challenges related to translating each AI ethical principle into practice. \citet{sanderson2023ai} carried out semi-structured interviews with AI designers and developers and discovered several challenges and insights related to translating each of Australia's AI Ethics Principles\footnotemark[\value{footnote}]. 

Studies have investigated the challenges faced by AI practitioners in incorporating specific ethical principles of AI. For instance, \citet{madaio2022assessing} conducted semi-structured interviews and workshops with 33 AI practitioners to investigate the challenges and needs of AI practitioners to assess the fairness of AI-based systems. \citet{holstein2019improving} conducted semi-structured interviews with 35 ML practitioners and a survey of 267 ML practitioners to investigate the team's challenge and needs in developing fairer ML systems. \citet{rakova2021responsible} conducted semi-structured interviews with AI practitioners with a focus on algorithmic accountability and investigated common challenges, ethical tensions, and effective enablers for responsible AI initiatives. On the other hand, studies have been conducted to explore the challenges related to addressing ethical issues of AI through focus groups with AI engineers \citep{christodoulou2021democracy}.

Most studies have either focused on investigating the challenges of AI practitioners in following specific AI ethical principles like fairness \citep{orr2020attributions, madaio2022assessing} and accountability \citep{rakova2021responsible}, explored the broader issues of addressing ethical concerns in AI \citep{christodoulou2021democracy} or their primary focus was not on investigating AI practitioners' challenges in incorporating AI ethics but have still managed to uncover certain challenges encountered by AI practitioners \citep{govia2020coproduction, vakkuri2019ethically}. It indicated that there is a lack of research that focuses primarily on investigating the \emph{challenges} of AI practitioners in incorporating ethics in AI. 

\section{Research Methodology} \label{Sec:Methodology}
We used survey research method to conduct this study to gather insights from broader AI practitioners on different ethics-related aspects of AI, such as their \emph{awareness} of AI ethics and \emph{challenges} to incorporate ethics in AI \citep{kasunic2005designing,wohlin2012experimentation}. 

\subsection{Survey Design}
\begin{figure}[htbp]
\centering
   \includegraphics [width=\linewidth] {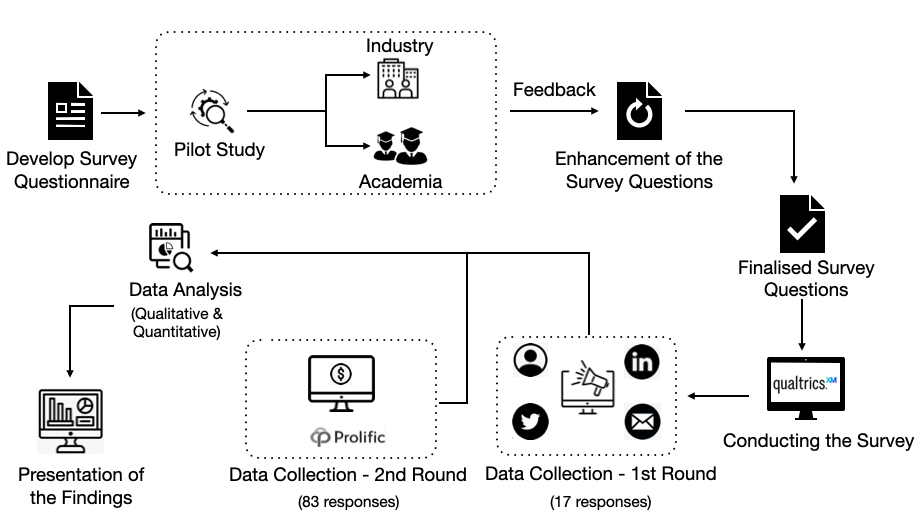}
\caption{Overview of the research methodology of the study.}
\label{fig:1}       
\end{figure}

We aimed at obtaining an overview of the participants' perspectives on two aspects of ethics (\emph{awareness} and \emph{challenges}) in AI based on their experiences through our survey (Appendix \ref{Appendix A}). Figure \ref{fig:1} shows the steps that we followed to design the survey to achieve the objective of this study. The survey planning was carried out from August 2022 to October 2022. During this phase, the main tasks performed were defining survey goals and variables, designing the questionnaire through iterative processes, and prioritising important survey questions. Hence, we designed the survey with both open and closed-ended questions. The survey was divided into three main sections and comprised 15 questions (12 closed-ended and 3 open-ended questions). The questions focused on several areas aligned with our future studies.

\subsubsection{Participant Information}
The first section of the survey was designed to collect \textbf{basic} demographic information and current employment details of the participants. The survey was anonymous and did not record any personally identifiable information, only information about their gender, age ranges, country of residence, and educational qualifications was collected. Employment details such as job roles/titles and primary responsibilities were collected to identify whether the participants are involved in AI-based software development activities. Participants were asked about their years of experience in AI-based software development to exclude those from the study data who had no experience in that field. All the participants included in our study have at least some experience in AI-based software development.

\subsubsection{Understanding the Participants' Perspectives of AI Ethics}
The second and third sections were designed to get an overview of the participants' perspectives on two aspects of ethics in AI including their \emph{awareness} of AI ethics and their \emph{challenges} in incorporating ethics in AI. In section two, we focused on identifying how aware AI practitioners are of the concept of AI ethics and AI ethical principles (we asked the participants to rate their familiarity with the concept of `ethics' in AI and report what ethical principles of AI they are aware of based on their experience), reasons for their awareness (we provided them with a list of reasons to become aware of AI ethics that were identified in our previous study \citep{pant2022AIethics} and asked them to provide any other reason(s) via open-ended option in the end), and their awareness of the role of formal education/ training (ranging from ``Extremely well" to ``Not at all") in preparing them to incorporate ethics during the process of developing AI-based systems. The significance of formal education in promoting the ethical implementation of AI has been discussed in the literature \citep{chivukula2020dimensions}.

In section three, we aimed to obtain the participants' perspectives on the challenges they encounter when incorporating ethics in AI. We used Australia's AI Ethics Principles\footnotemark[\value{footnote}] as a reference and requested participants to rate the degree of challenges (``Very Challenging'' to ``No experience'') they experienced in adhering to the ethical principles while developing AI-based systems.
Moreover, we included an open-ended question to gather the participants' opinions on the primary challenges or barriers they encounter in incorporating ethics in AI. This question enabled us to gain a range of viewpoints on the challenges faced by AI practitioners in incorporating AI ethics.

\subsubsection{Pilot Study}
Once the survey had been formulated, a preliminary trial was carried out with AI practitioners from our network in order to confirm the questions' clarity and comprehensibility, gauge the amount of time needed to complete the survey, and gather their opinions on ways to enhance it. The survey was subsequently circulated among two AI practitioners engaged in AI development activities and two academics with previous involvement in such activities. Taking their suggestions into account, we made some modifications to the survey questions and added some definitions to enhance the clarity of some terms. After incorporating these changes, we finalised the survey and conducted our study.

\subsection{Survey Sampling and Data Collection}
We adopted a non-probability purposive sampling technique for our study, which entails selecting participants based on particular characteristics rather than their availability \citep{baltes2022sampling}. This method allowed us to specifically target our desired group of participants, who are AI practitioners engaged in AI development activities. Our survey was created using the \emph{Qualtrics} platform and we advertised the survey as an anonymous survey link after obtaining the required ethics approval (Reference Number: 34685). The survey questionnaire can be found in Appendix \ref{Appendix A}.

We carried out two rounds of data collection. The first round of data collection was executed from October 2022 to December 2022 and we advertised the survey on social media groups like LinkedIn and Twitter, targeting AI practitioners who are engaged in AI development. We received complete survey responses from 17 AI practitioners who have some level of experience in AI development. Since we didn't receive enough responses in the initial data collection, we decided to carry out a second round. We used the Prolific platform from January 2023 to February 2023 to advertise the survey. As a result, we obtained valuable information from 83 AI practitioners who participated in the survey. The Prolific platform has useful features to filter and select participants, as well as customise monetary incentives for each of the participants. As our aim was to obtain responses from participants who had some level of experience in AI development activities, we employed the participant filtering options of ``Employment Sector - Information Technology'' and ``Employment Status - Full-Time/Part-Time''. We provided the participants who completed the survey through Prolific with a reward of 13.30 AUD. Since the survey was shared globally, we obtained responses from various countries, as illustrated in Table \ref{tab:demo}. As we did not have any particular preference for certain countries, the responses were distributed across various regions. We obtained a majority of the responses from Europe (30\%), followed by the participants from other continents like Africa (28\%), North America (18\%), Asia (12\%), etc. Section \ref{sec:demo} contains an in-depth analysis of the participants' demographics.

\begin{table}[htbp]
\centering
\caption{Data sources and analysis types used to answer RQs (descriptive statistics for quantitative data analysis and STGT for qualitative data analysis).} \label{table:1}     

\scriptsize
\begin{tabular} {>{\raggedright\arraybackslash}p{0.5cm}>{\raggedright\arraybackslash}p{2.2cm}>{\raggedright\arraybackslash}p{2.2cm}>{\raggedright\arraybackslash}p{5.5cm}}

 
\hline\noalign{\smallskip}
RQ & Data Source & Data Analysis Type & Purpose of Analysis\\
\hline\noalign{\smallskip}
RQ1 & Closed-ended question & Quantitative analysis & To get an overview of participants' awareness of different aspects related to AI ethics including (i) extent of awareness of ‘AI ethics’ concept, (ii) awareness of AI ethical principles, (iii) reasons for awareness, and (iv) role of formal education/training in preparing AI practitioners to incorporate AI ethics\\

RQ2 &  Closed-ended, Follow-up open-ended question & Quantitative analysis Qualitative analysis  & To get an overview of the extent to which AI practitioners are challenged to consider and follow each AI ethical principle and the key challenges they face in incorporating AI ethics \\

\noalign{\smallskip}\hline
\end{tabular}
\end{table}

\subsection{Data Analysis} \label{data analysis}
\begin{figure}[htbp]
\centering
   \includegraphics [width=\linewidth] {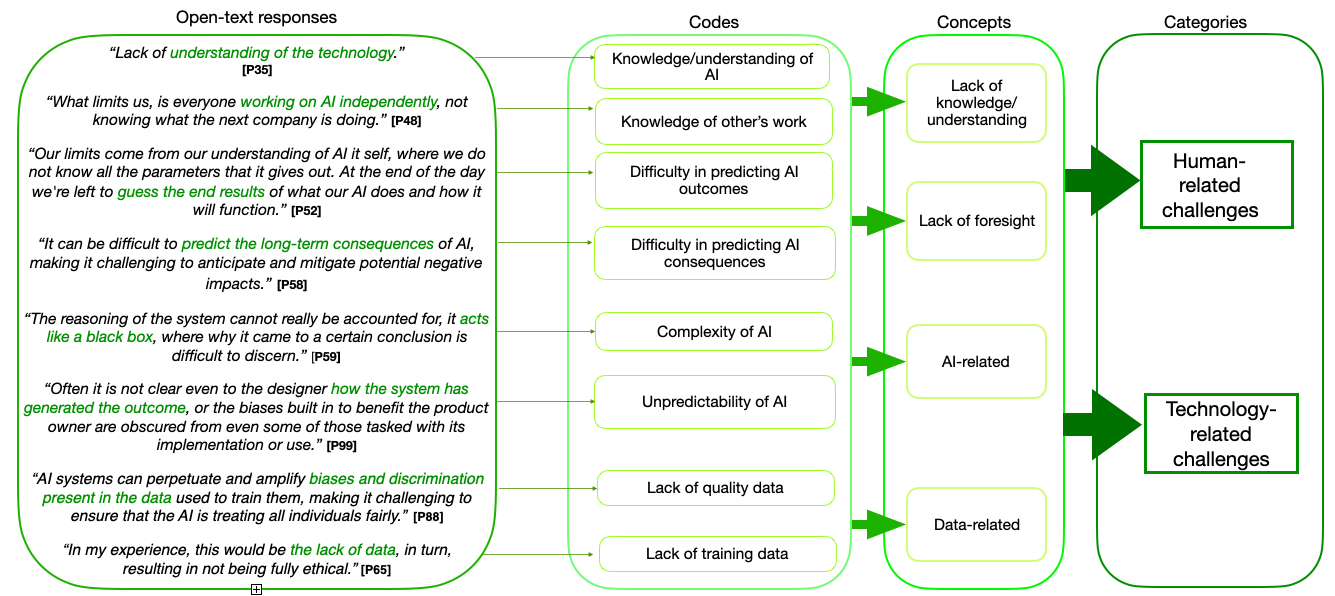}
\caption{Examples of STGT analysis \citep{hoda2021socio} applied to qualitative data on challenges/barriers in incorporating ethics in AI.}
\label{fig:HumanLim}       
\end{figure}

We collected both qualitative and quantitative data through our survey. So, we used a mixed-method approach to analyse the survey data. The data analysis types (quantitative/qualitative) that we used to address each research question in this study are presented in Table \ref{table:1}. We used Microsoft Excel to statistically analyse the quantitative data and organise the qualitative data. Meanwhile, we used the \emph{Socio-Technical Grounded Theory (STGT) for Data Analysis} method to analyse the qualitative data \citep{hoda2021socio}, which is well-suited for analysing qualitative data, such as those acquired from open-ended text-based survey responses (or \emph{open-text} for short). The purpose of the \emph{STGT for data analysis}, unlike a complete STGT study, is not to develop advanced theories, but rather to identify important patterns in the qualitative data and present them as layered and/or multi-dimensional findings, along with insights and reflections. To do this, we used an \emph{open coding} approach to developing concepts and categories with \emph{constant comparison} of various open-text responses. This approach does not require extensive qualitative data. For example, we gathered open-text responses from 100 participants for the question, \emph{``In your experience, what are the main challenges or barriers in incorporating ethics in AI?"}. These responses can be analysed using the  \emph{STGT for data analysis} approach. We applied open coding in open-text answers and developed codes as shown in Figure \ref{fig:HumanLim}. Codes such as \emph{`lack of knowledge/ understanding of AI'}, \emph{`lack of knowledge of other's work'} were grouped together to create the concept \emph{`lack of knowledge/ understanding'}. Likewise, codes such as \emph{`difficulty in predicting AI outcomes'} and \emph{`difficulty in predicting AI consequences'} were grouped together to create the concept \emph{`lack of foresight'}.

For example, in regard to the question mentioned before, we identified the set of concepts as various types of challenges in incorporating ethics in AI. As a result, we grouped together all the concepts that are related to humans to form a category called \emph{`human-related challenges'}. On the other hand, we derived multiple codes which were related to the challenges of technology, as presented in Figure \ref{fig:HumanLim}. This led to the formation of the other high-level category, namely: \emph{`technology-related challenges'}. In this way, we derived a total of three categories of challenges including \textbf{general challenges}, \textbf{technology-related challenges}, and \textbf{human-related challenges}. Detailed information on these challenges is provided in Section \ref{Findings}.

The survey questionnaire was designed by four authors, and all five authors participated in analysing the data and presenting the findings. We engaged in multiple rounds of discussion for each stage to reach decisions. During the analysis phase, the first author analysed the quantitative data and shared the results with the other authors. They discussed the optimal approach to presenting the findings. Additionally, in the analysis of the qualitative data, the second and third authors assisted the first author. Once the qualitative data were analysed, the results, including codes, concepts, and categories for the open-ended question, were shared and discussed among all authors who also helped in presenting the findings.

Open coding, constant comparison, and memoing are the steps involved in the STGT method for data analysis. \emph{``Basic memoing is the process of documenting the researcher’s thoughts, ideas, and reflections on emerging concepts and (sub)categories and evidence-based conjectures on possible links between them"} \citep{hoda2021socio}. Therefore, we created memos to document important insights and reflections that we discovered during the process of open coding activities. An example of a memo that we created for the concept \emph{`lack of knowledge/understanding'} is shown below. The discussion on the key insights from memoing is presented in Section \ref{sec:insights}.

\begin{tikzpicture} 
\centering
\node[draw,align=left] at (3,0) { \textbf{Memo on ``Lack of knowledge/understanding":} ``Not enough details here to\\ gather what kind  of knowledge is lacking. Some shared include lack of \\ knowledge of AI outputs, ethical implications of AI, and lack of programming \\skills (although unclear what this means for incorporating ethics - or just that\\ AI implementation is becoming very complex) \\
Link: P97 and P36 mention limitations of human knowledge (about limits of AI\\ - P36); limitations of humans to control and alter AI decisions (P41) and link it\\ to possible misuse of ethical loopholes by AI (see "powerful and inescapable AI"\\ memo). P48 notes a lack of human overview of what others are doing and what\\ we are creating (no overview and coordination). P58 notes a lack of human\\ ability to predict the long-term consequences of AI (no long-term view)."
};
\end{tikzpicture}

\section{Findings} \label{Findings}

\subsection{Participants' Demographics} \label{sec:demo}
In this section, we present the demographics of the survey participants. Table \ref{tab:demo} summarises the overall statistics of the participants based on their gender, age, country, work experience, education, and job title. 

A total of 100 AI practitioners participated in the survey. The majority of the participants were male (71\%) whereas only 29\% were females. The most common age group of the participants was 26 to 30 years, comprising 31\% of the sample followed by the age group ranging from 20-25 years (25\%). Only 3\% of the participants were over the age of 50. Similarly, the majority of the participants (23\%) were AI/data analysts followed by AI developers (19\%). 16\% of the participants were AI engineers followed by 11\% participants who were AI/ML practitioners. The job title of only 3\% of the participants was AI/ML specialist whereas 14\% of the participants were in the `Other' group. Their job titles included software engineer or software developer. As our target survey participants were AI practitioners involved in AI development activities, we wanted to know the major AI development-related activities they were involved in. The results show that the majority of the participants (19.1\%) were involved in the \emph{Data Collection} activity followed by the \emph{Data Cleaning} activity (15.6\%).

\begin{table}[htbp]
\caption{\small Demographics of the Survey Participants.}
\label{tab:demo}
\resizebox{\columnwidth}{!}{
\scriptsize
\begin{tabular}{@{}llll@{}}
\toprule
Gender &  & Work Experience &  \\ \midrule
Men     & \mybarhhigh{15.72}{71\%}                         & Less than 1 year      & \mybarhhigh{20.49}{28\%}\\
Women   & \mybarhhigh{4.04}{29\%}                         & 1 to 2 years         & \mybarhhigh{35.72}{43\%}\\
   &                           & 3 to 5 years         & \mybarhhigh{23.59}{19\%}\\
\cmidrule(r){1-1}
\textbf{Age}   &          & 6 to 10 years         & \mybarhhigh{11.23}{7\%}\\
\cmidrule(r){1-1}
20 to 25 years     & \mybarhhigh{21.1}{25\%}              & 11 to 15 years        & \mybarhhigh{5.35}{3\%}\\
\cmidrule(r){3-3}
26 to 30 years     & \mybarhhigh{27.2}{31\%}              &  \textbf{Education}        & \\
\cmidrule(r){3-3}
31 to 35 years     & \mybarhhigh{16.5}{17\%}              & High school      & \mybarhhigh{11.23}{10\%}  \\
36 to 40 years     & \mybarhhigh{15.86}{16\%}               & Bachelor Degree      & \mybarhhigh{42.23}{56\%} \\
41 to 45 years     & \mybarhhigh{7.37}{6\%}                    &  Master Degree    & \mybarhhigh{21.23}{22\%} \\
46 to 50 years     &  \mybarhhigh{3.37}{2\%}                  &  Ph.D. or higher       & \mybarhhigh{11.23}{7\%} \\
Above 50 years     &  \mybarhhigh{4.86}{3\%}                      &  Prefer not to answer    & \mybarhhigh{2.}{1\%} \\
\cmidrule(r){1-1}
\textbf{Job Title} &    &  Others & \mybarhhigh{6.23}{4\%} \\
\cmidrule(r){1-1}
\cmidrule(r){3-3}
AI Expert & \mybarhhigh{7.37}{6\%} & \textbf{Countries} & \\
\cmidrule(r){3-3}
AI/ML Specialist & \mybarhhigh{4.86}{3\%} &   Africa  &  \mybarhhigh{25.23}{28\%}    \\
AI/Data Scientist & \mybarhhigh{20}{23\%} &  North America   &  \mybarhhigh{16.95}{18\%} \\
AI Designer & \mybarhhigh{8.37}{8\%} & Europe & \mybarhhigh{27.23}{30\%}  \\
AI/ML Practitioner & \mybarhhigh{10.37}{11\%} & South America    & \mybarhhigh{5.73}{3\%} \\
AI Engineer & \mybarhhigh{15}{16\%} & Oceania           &  \mybarhhigh{8.23}{9\%} \\
AI Developer & \mybarhhigh{17.37}{19\%} & Asia & \mybarhhigh{12.23}{12\%} \\
Others & \mybarhhigh{13.40}{14\%}
\\

\bottomrule
\end{tabular}
}
\end{table}

\subsection{RQ1 -- How aware are AI practitioners of different aspects related to AI ethics?} \label{RQ1}
\subsubsection{Extent of awareness of `AI ethics' concept} \label{FindingsAwarenessethics}
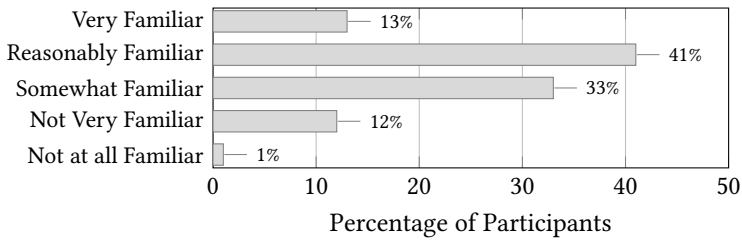
\begin{figure*}[ht]
\centering
\begin{tikzpicture}
\begin{axis}[%
xlabel= Percentage of Participants,
width=0.70\textwidth, height=1.45in,
xbar, bar width=8pt,
xmin=0, xmax=50,
symbolic y coords={Not at all Familiar, Not Very Familiar, Somewhat Familiar, Reasonably Familiar, Very Familiar},
ytick=\empty,
xmajorgrids
]
\addplot [fill=gray!30,draw=gray] coordinates {(1,Not at all Familiar) (12,Not Very Familiar) (33,Somewhat Familiar) (41,Reasonably Familiar) (13,Very Familiar)}
node [minimum size=0pt,inner sep=0pt,right,pin={[font=\footnotesize,pin distance=0.3cm]0:1\% }] at (axis cs:1,Not at all Familiar) {}
node [minimum size=0pt,inner sep=0pt,right,pin={[font=\footnotesize,pin distance=0.3cm]0:12\% }] at (axis cs:12,Not Very Familiar) {}
node [minimum size=0pt,inner sep=0pt,right,pin={[font=\footnotesize,pin distance=0.3cm]0:33\% }] at (axis cs:33,Somewhat Familiar) {}
node [minimum size=0pt,inner sep=0pt,right,pin={[font=\footnotesize,pin distance=0.3cm]0:41\% }] at (axis cs:41,Reasonably Familiar) {}
node [minimum size=0pt,inner sep=0pt,right,pin={[font=\footnotesize,pin distance=0.3cm]0:13\%}] at (axis cs:13,Very Familiar) {}
coordinate (symbh) at (axis cs:0,Not at all Familiar)
coordinate (symb) at (axis cs:0,Not Very Familiar)
coordinate (num) at (axis cs:0,Somewhat Familiar)
coordinate (uprcase) at (axis cs:0,Reasonably Familiar)
coordinate (lwrcase) at (axis cs:0,Very Familiar);
\end{axis}
\node [font=\small, left] at (symbh) {Not at all Familiar};
\node [font=\small, left] at (symb) {Not Very Familiar};
\node [font=\small,left] at (num) {Somewhat Familiar};
\node [font=\small,left] at (uprcase) {Reasonably Familiar};
\node [font=\small,left] at (lwrcase) {Very Familiar};
\end{tikzpicture}
\caption{AI practitioner’s degree of familiarity with the concept of AI ethics.}
\label{fig:AwarenessEthics}
\end{figure*}

In our previous work (GTLR) \citep{pant2022AIethics}, we identified five categories discussing AI practitioners’ viewpoints on ethics in AI among which \emph{`developer awareness'} is one of them. Under this category, there are multiple concepts, and developer awareness of \emph{`AI ethics and ethical principles'} is one of the underlining concepts. The main reason to conduct the GTLR (our previous work) over a Systematic Literature Review (SLR) was due to the lack of empirical studies focusing on investigating AI practitioners' views and understanding of ethics in AI. Therefore, getting an \textbf{industry} perspective on this concept by asking them about their level of familiarity with the concept of ethics when it relates to AI development was required. The participants were given five different levels of familiarity to choose from, ranging from `Very familiar' to `Not at all familiar.' As shown in Figure \ref{fig:AwarenessEthics}, out of all participants, 41\% had a reasonable familiarity with the concept of AI ethics, followed by 33\% who were somewhat familiar with the concept. Conversely, only 13 \% of the participants had a high level of familiarity with ethics in AI, 12\% of the participants had a low level of familiarity and only 1\% of the participants having no familiarity with the concept at all. 

\subsubsection{Awareness of AI ethical principles} \label{FindingsAwarenessethprinciple}
The ethical guidelines for AI differ depending on the country and organisation. Our previous work (GTLR) revealed that AI practitioners discussed only four specific ethical principles of AI namely \emph{`Accountability'}, \emph{`Fairness'}, \emph{`Transparency and explainability'}, and \emph{`Privacy'}. However, obtaining a broader perspective from the industry on all the ethical principles related to AI that practitioners were aware of was required. To achieve this, we referred to Australia's AI Ethics Principles\footnotemark[\value{footnote}] as a basis for investigation.

\begin{figure*}[ht]
\centering
\begin{tikzpicture}
\begin{axis}[%
xlabel= Percentage of Participants,
width=0.70\textwidth, height=2.1in,
xbar, bar width=8pt,
xmin=0, xmax=50,
symbolic y coords={None,All,Contestability,Human societal and env. wellbeing,Transparency and explainability,Fairness,Accountability,Human-centred values,Reliability and safety,Privacy protection and security},
ytick=\empty,
xmajorgrids
]
\addplot [fill=gray!30,draw=gray] coordinates {(0.56,None) (3.38,All) (3.38,Contestability) (8.17,Human societal and env. wellbeing) (11.83,Transparency and explainability) (12.11,Fairness) (12.68,Accountability) (14.93,Human-centred values) (14.93,Reliability and safety) (18.03,Privacy protection and security)}
node [minimum size=0pt,inner sep=0pt,right,pin={[font=\footnotesize,pin distance=0.3cm]0:0.56\%}] at (axis cs:0.56,None) {}
node [minimum size=0pt,inner sep=0pt,right,pin={[font=\footnotesize,pin distance=0.3cm]0:3.38\%}] at (axis cs:3.38,All) {}
node [minimum size=0pt,inner sep=0pt,right,pin={[font=\footnotesize,pin distance=0.3cm]0:3.38\%}] at (axis cs:3.38,Contestability) {}
node [minimum size=0pt,inner sep=0pt,right,pin={[font=\footnotesize,pin distance=0.3cm]0:8.17\%}] at (axis cs:8.17,Human societal and env. wellbeing) {}
node [minimum size=0pt,inner sep=0pt,right,pin={[font=\footnotesize,pin distance=0.3cm]0:11.83\%}] at (axis cs:11.83,Transparency and explainability) {}
node [minimum size=0pt,inner sep=0pt,right,pin={[font=\footnotesize,pin distance=0.3cm]0:12.11\%}] at (axis cs:12.11,Fairness) {}
node [minimum size=0pt,inner sep=0pt,right,pin={[font=\footnotesize,pin distance=0.3cm]0:12.68\%}] at (axis cs:12.68,Accountability) {}
node [minimum size=0pt,inner sep=0pt,right,pin={[font=\footnotesize,pin distance=0.3cm]0:14.93\%}] at (axis cs:14.93,Human-centred values) {}
node [minimum size=0pt,inner sep=0pt,right,pin={[font=\footnotesize,pin distance=0.3cm]0:14.93\%}] at (axis cs:14.93,Reliability and safety) {}
node [minimum size=0pt,inner sep=0pt,right,pin={[font=\footnotesize,pin distance=0.3cm]0:18.03\%}] at (axis cs:18.03,Privacy protection and security) {}
coordinate (fy) at (axis cs:0,None)
coordinate (gy) at (axis cs:0,All)
coordinate (dy) at (axis cs:0,Contestability)
coordinate (ay) at (axis cs:0,Human societal and env. wellbeing)
coordinate (cy) at (axis cs:0,Transparency and explainability)
coordinate (hy) at (axis cs:0,Fairness)
coordinate (ey) at (axis cs:0,Accountability)
coordinate (by) at (axis cs:0,Human-centred values)
coordinate (iy) at (axis cs:0,Reliability and safety)
coordinate (jy) at (axis cs:0,Privacy protection and security);
\end{axis}
\node [font=\small,left] at (fy) {None};
\node [font=\small,left] at (gy) {All};
\node [font=\small,left] at (dy) {Contestability};
\node [font=\small,left] at (ay) {Human societal and env. wellbeing};
\node [font=\small,left] at (cy) {Transparency and explainability};
\node [font=\small,left] at (hy) {Fairness};
\node [font=\small,left] at (ey) {Accountability};
\node [font=\small,left] at (by) {Human-centred values};
\node [font=\small,left] at (iy) {Reliability and safety};
\node [font=\small,left] at (jy) {Privacy protection and security};
\end{tikzpicture}
\caption{AI practitioner’s awareness of AI ethical principles.}
\label{AwarenessEthPrinciple}
\end{figure*}
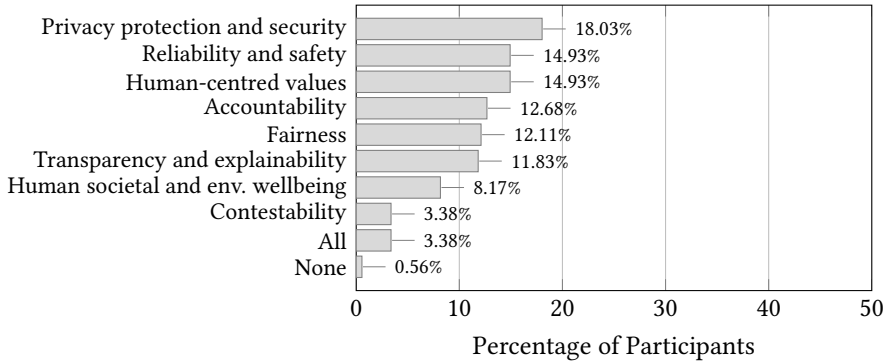

We found that the majority of the participants were aware of the AI ethical principle namely, \emph{`Privacy protection and security'} (18.03\%) followed by two ethical principles, \emph{`Reliability and safety'} and \emph{`Human-centred values'} with 14.93\% each as shown in Figure \ref{AwarenessEthPrinciple}. Additionally, most of the participants were also aware of the three AI ethical principles including \emph{`Accountability'} (12.68\%) followed by \emph{`Fairness'} (12.11\%) and \emph{`Transparency and explainability'} (11.83\%). Only a small percentage of the participants (3.38\%) were aware of all AI ethical principles whereas 0.56\% were not familiar with any of the ethical principles of AI. 

\subsubsection{Reasons for awareness} \label{sec: reasons}
Several studies have reported the reasons that contribute to the awareness of AI ethics and ethical principles among AI practitioners such as exposure to news and media, customer complaints \citep{holstein2019improving}, personal interests, and experiences \citep{ibanez2022operationalising}. Therefore, we aimed to conduct further research on these reasons and also explore other factors that raise the awareness of AI practitioners regarding AI ethics and ethical principles. To achieve this, we presented a list of possible reasons and asked participants to choose the applicable ones and add any other reasons that were not listed in the options through an open-ended answer.

\begin{figure*}[ht]
\centering
\begin{tikzpicture}
\begin{axis}[%
xlabel= Percentage of Participants,
width=0.70\textwidth, height=1.70in,
xbar, bar width=8pt,
xmin=0, xmax=50,
symbolic y coords={Others, Customer complaints, First hand professional experience, First hand personal experience, I have a personal interest, Through news and media, Workplace rules and policies},
ytick=\empty,
xmajorgrids
]
\addplot [fill=gray!30,draw=gray] coordinates {(0.74,Others) (6.25,Customer complaints) (14.34,First hand professional experience) (17.65,First hand personal experience) (18.01,I have a personal interest) (19.85,Through news and media) (23.16,Workplace rules and policies)}
node [minimum size=0pt,inner sep=0pt,right,pin={[font=\footnotesize,pin distance=0.3cm]0:0.74\%}] at (axis cs:0.74,Others) {}
node [minimum size=0pt,inner sep=0pt,right,pin={[font=\footnotesize,pin distance=0.3cm]0:6.25\% }] at (axis cs:6.25,Customer complaints) {}
node [minimum size=0pt,inner sep=0pt,right,pin={[font=\footnotesize,pin distance=0.3cm]0:14.34\% }] at (axis cs:14.34,First hand professional experience) {}
node [minimum size=0pt,inner sep=0pt,right,pin={[font=\footnotesize,pin distance=0.3cm]0:17.65\% }] at (axis cs:17.65,First hand personal experience) {}
node [minimum size=0pt,inner sep=0pt,right,pin={[font=\footnotesize,pin distance=0.3cm]0:18.01\%}] at (axis cs:18.01,I have a personal interest) {}
node [minimum size=0pt,inner sep=0pt,right,pin={[font=\footnotesize,pin distance=0.3cm]0:19.85\%}] at (axis cs:19.85,Through news and media) {}
node [minimum size=0pt,inner sep=0pt,right,pin={[font=\footnotesize,pin distance=0.3cm]0:23.16\% }] at (axis cs:23.16,Workplace rules and policies) {}
coordinate (symbh) at (axis cs:0,Others)
coordinate (symb) at (axis cs:0,Customer complaints)
coordinate (num) at (axis cs:0,First hand professional experience)
coordinate (uprcase) at (axis cs:0,First hand personal experience)
coordinate (lwrcase) at (axis cs:0,I have a personal interest)
coordinate (lwrcasew) at (axis cs:0,Through news and media)
coordinate (lwrcasewq) at (axis cs:0,Workplace rules and policies);
\end{axis}
\node [font=\small, left] at (symbh) {Others};
\node [font=\small, left] at (symb) {Customer complaints};
\node [font=\small,left] at (num) {First hand professional experience};
\node [font=\small,left] at (uprcase) {First hand personal experience};
\node [font=\small,left] at (lwrcase) {I have a personal interest};
\node [font=\small,left] at (lwrcasew) {Through news and media};
\node [font=\small,left] at (lwrcasewq) {Workplace rules and policies};
\end{tikzpicture}
\caption{AI practitioner's reasons for awareness of ethics in AI}
\label{AwarenessReason}
\end{figure*}
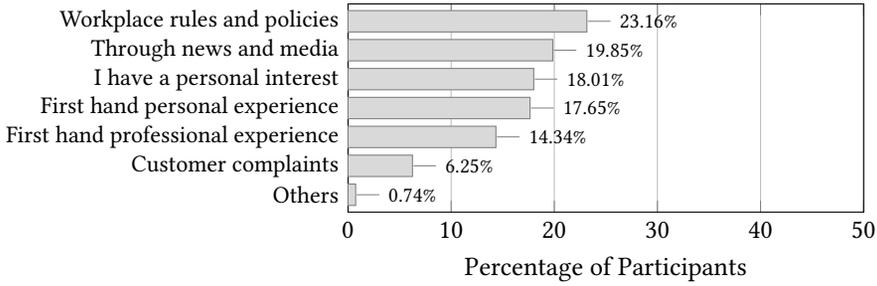

According to our findings, the reason for the majority of the AI practitioners' awareness of AI ethics and ethical principles was \emph{workplace rules and policies}, accounting for 23.16\% of responses, as shown in Figure \ref{AwarenessReason}. The second reason, cited by 19.85\% of participants, was awareness of AI ethics and ethical principles through \emph{news and media}. Likewise, the personal interests of the AI practitioners and their first-hand personal experience were also the reasons for their awareness of AI ethics and ethical principles that accounted for 18.01\% and 17.65\% of responses respectively. A mere 0.74\% of the participants provided explanations for their reasons for awareness of AI ethics and ethical principles through the open-ended answer option. Interestingly, all participants who provided an answer cited ``University" as their reason for awareness. For instance, participant [P38] explicitly stated ``University courses" as their source, while [P63] simply mentioned ``University".

\subsubsection{Role of formal education/training in preparing AI practitioners to incorporate AI ethics} \label{Findings_formaleducation}
 
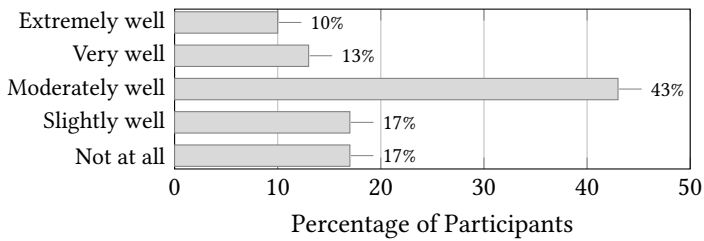
\begin{figure*}[ht]
\centering
\begin{tikzpicture}
\begin{axis}[
xlabel= Percentage of Participants,
width=0.70\textwidth, height=1.45in,
xbar, bar width=8pt,
xmin=0, xmax=50,
symbolic y coords={Not at all, Slightly well, Moderately well, Very well, Extremely well},
ytick=\empty,
xmajorgrids
]
\addplot [fill=gray!30,draw=gray] coordinates {(17,Not at all) (17,Slightly well) (43,Moderately well) (13,Very well) (10,Extremely well)}
node [minimum size=0pt,inner sep=0pt,right,pin={[font=\footnotesize,pin distance=0.3cm]0:17\%}] at (axis cs:17,Not at all) {}
node [minimum size=0pt,inner sep=0pt,right,pin={[font=\footnotesize,pin distance=0.3cm]0:17\%}] at (axis cs:17,Slightly well) {}
node [minimum size=0pt,inner sep=0pt,right,pin={[font=\footnotesize,pin distance=0.3cm]0:43\%}] at (axis cs:43,Moderately well) {}
node [minimum size=0pt,inner sep=0pt,right,pin={[font=\footnotesize,pin distance=0.3cm]0:13\%}] at (axis cs:13,Very well) {}
node [minimum size=0pt,inner sep=0pt,right,pin={[font=\footnotesize,pin distance=0.3cm]0:10\%}] at (axis cs:10,Extremely well) {}
coordinate (symbh) at (axis cs:0,Not at all)
coordinate (symb) at (axis cs:0,Slightly well)
coordinate (num) at (axis cs:0,Moderately well)
coordinate (uprcase) at (axis cs:0,Very well)
coordinate (lwrcase) at (axis cs:0,Extremely well);
\end{axis}
\node [font=\small, left] at (symbh) {Not at all};
\node [font=\small, left] at (symb) {Slightly well};
\node [font=\small,left] at (num) {Moderately well};
\node [font=\small,left] at (uprcase) {Very well};
\node [font=\small,left] at (lwrcase) {Extremely well};
\end{tikzpicture}
\caption{Role of formal education/training in preparing AI practitioners to incorporate AI ethics}
\label{fig:Awareness_Education}
\end{figure*}

The significance of educating people about ethics in AI has been discussed in the literature \citep{chivukula2020dimensions} which was one of the recommendations in our previous work (GTLR) \citep{pant2022AIethics}. Therefore, our next objective was to gather AI practitioners' perspectives on how effective formal education or training helps them in preparing themselves to incorporate ethics during AI development. To accomplish this, participants were asked to rate the extent to which formal education or training help them, using a five-point scale ranging from `Extremely well' to `Not at all.' Figure \ref{fig:Awareness_Education} demonstrates that the majority of participants (43\%) felt that formal education or training moderately help them to incorporate ethics in AI. Conversely, 17\% of the participants stated that formal education or training provides slight assistance while another 17\% reported that it provides no assistance at all. It is worth noting that only a small number of participants (10\%) believed that formal education or training is extremely effective in preparing them in incorporating ethics in AI.

\subsection{RQ2 --  What challenges/barriers do AI practitioners face in incorporating ethics in
AI?} \label{Findings_RQ2}
First, we began by asking a closed-ended question to the participants to evaluate the ethical principles of AI by assessing the challenges involved in considering and adhering to each one. We used Australia’s AI Ethics Principles\footnotemark[\value{footnote}] as a reference and requested the participants to rate the degree of difficulty (ranging from ``Very Challenging” to ``No experience”).

The results indicate that the majority of the participants (27\%) found \emph{`Human-centred values'} most challenging to adhere to while developing AI-based systems, followed by another principle \emph{`Privacy protection and security'} (26\%). In addition, 24\% and 22\% of the participants found \emph{`Transparency and explainability'} and \emph{`Reliability and safety'} very challenging principles respectively to adhere to. Besides, the majority of the participants (12\%) also mentioned that \emph{`Privacy protection and security'} is the least challenging ethical principle to adhere to during AI development. Only 3\% of the participants reported that \emph{`Human-centred values'} is the ethical principle that is least challenging to adhere to during AI development. 
 \begin{figure}
\centering
\begin{tikzpicture}
\begin{axis}[
    ybar,
    width=\textwidth,
    height=1.8in,
    font=\scriptsize,
    major x tick style = transparent,
    ybar=2*\pgflinewidth,
    bar width=3pt,
    xlabel={AI Ethical Principle},
    ylabel={Percentage of Participants},
    xtick=data,
    scaled y ticks = false,
    xticklabel style={
        align=center 
    },
    xticklabels={
        {Account-\\ability}, 
        {Contest-\\ability},
        {Fairness},
        {Human-\\centred\\ values},
        {Human, \\societal \& \\env. wellbeing},
        {Privacy \\protection \& \\security},
        {Reliability \\\& \\ safety},
        {Transparency \\\& \\ explainability}
        },
    enlarge x limits=0.075,
    ymin=0,
    legend style={
        at={(0.5,-0.6)},
        anchor=north,
        legend columns=-1,
        font=\scriptsize
    }
]
\addplot[fill=darkgray] coordinates {(1,17) (2,13) (3,19) (4,27) (5,19) (6,26) (7,22) (8,24)};
\addplot[fill=lightgray] coordinates {(1,21) (2,30) (3,23) (4,34) (5, 30) (6, 23) (7,32) (8,32)};
\addplot[fill=white] coordinates {(1,32) (2,26) (3,28) (4,20) (5, 27) (6, 16) (7,25) (8,23)};
\addplot[fill=brown] coordinates {(1,19) (2,15) (3,16) (4,12) (5, 15) (6, 21) (7,12) (8,6)};
\addplot[fill=black] coordinates {(1,5) (2,6) (3,8) (4,3) (5, 4) (6, 12) (7,5) (8,11)};
\addplot[fill=orange] coordinates {(1,6) (2,10) (3,6) (4,4) (5, 5) (6, 2) (7,4) (8,4)};

\legend{Very, Reasonably, Somewhat, Not very, Not at all, No experience}
\end{axis}
\end{tikzpicture}
\caption{AI practitioner’s degree of challenges in considering and following AI ethical principles.}
\label{challengesDegree}
\end{figure}
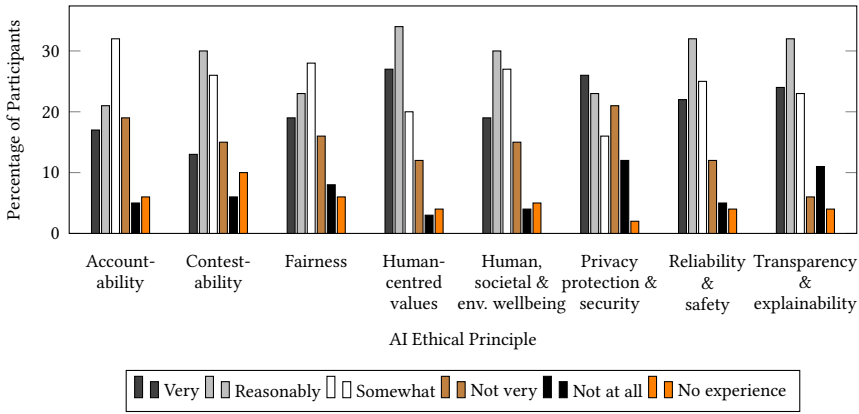

\begin{figure}[ht]
    \centering
    \includegraphics [width= \linewidth]{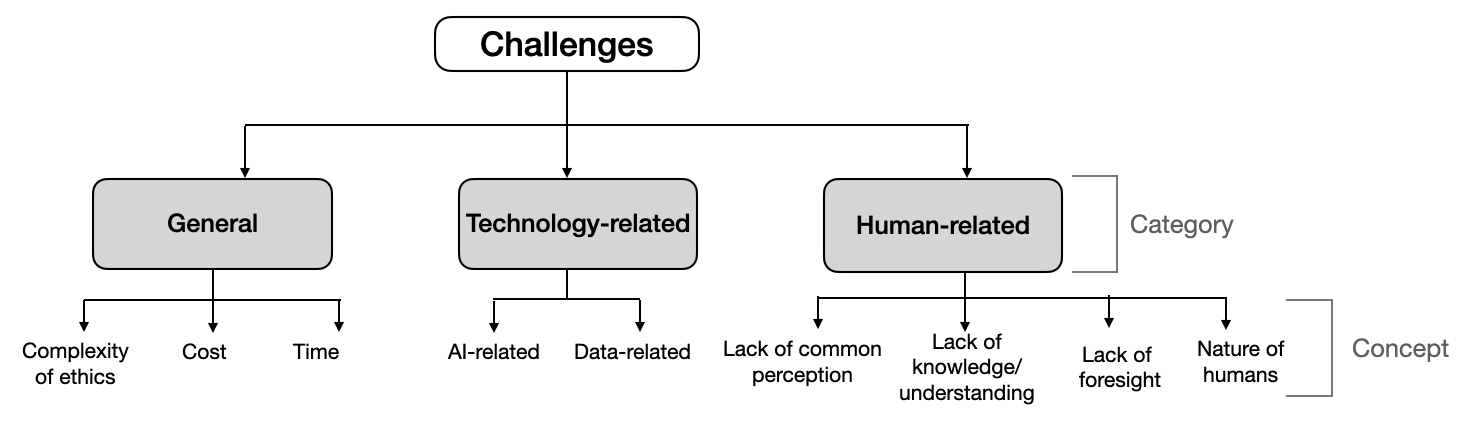}
    \caption{Overview of the challenges/barriers in incorporating ethics in AI}
    \label{Challenges_Overview}
\end{figure}

We asked an open-ended question to explore the main challenges or barriers that AI practitioners face to incorporate ethics in AI, and the qualitative data was analysed using the \emph{STGT method for data analysis}. Based on the response to the open-ended question, we categorised the main challenges of AI practitioners into three categories including (i) General challenges, (ii) Technology-related challenges, and (iii) Human-related challenges. These are explained in detail below. Figure \ref{Challenges_Overview} shows the overview of the challenges/barriers in incorporating ethics in AI that we obtained through the analysis of the qualitative data.

\subsubsection{General Challenges} \label{sec:General}
\subsubsection*{\textbf{Complexity of ethics}}
Quantification: The complexity of ethics arises from the multifaceted nature of ethical values and principles and makes it challenging to quantify, such as capturing, defining, and measuring them. Participants also identified this as one of the challenges in integrating ethics while developing AI-based systems. For example,
participants [P72] and [P37] said: \\
\faCommenting \hspace{0.05cm} \emph{``It is similar to values such as happiness, satisfaction, and quality of life. How to define them mathematically? In such a situation, the creators of the algorithm will probably reach for an economic, tangible, and imaginative argument - how much can be saved thanks to it. In addition, there are many exceptions to the rule that cannot be captured in the algorithm."} -- [P72] \\
\faCommenting \hspace{0.05cm} \emph{``Measuring and building ethics red lines is challenging"} -- [P37] \\

Translation: Translating ethics from principles to practice can be a challenging task. While ethical principles provide a foundation for ethical decision-making, applying these principles in practice can be complex. It was considered a challenge to incorporate ethical principles into AI-based systems for most of the participants (P2, P4, P5, P6, P7, P8, P10, P11, P12, P14, P15, P16, P17, P18, P19, P21, P23, P24, P35). For example, \\
\faCommenting \hspace{0.05cm} \emph{``The main challenge is to implement some huge concept onto program written in the programming language."} -- [P89] \\

Management: Various ethical principles for AI exist and differ among countries and organisations. Ensuring that these principles are balanced, connected, and monitored during the development of AI-based systems can present a challenge for AI practitioners. Some participants expressed similar views on the management of AI ethical principles and discussed the challenges they faced during AI development. \\
\faCommenting \hspace{0.05cm} \emph{``One has to decide how to connect it with the other principles and rules we base our AI upon in order to maximise our target outcomes. Lastly, since these things are very difficult to control and audit, there is no real incentive to introduce them into the system. "} -- [P30] \\
\faCommenting \hspace{0.05cm} \emph{``A balance on all of them seems difficult as is always on the software you can't have them all."} -- [P32] \\

Implementation: As mentioned earlier, there are several ethical principles of AI, and AI practitioners need to apply all of them to develop an ethical AI-based system. Following each ethical principle during  the development of an AI-based system could be a challenge as discussed by some of the participants (P32, P66, P71, P83).\\
\faCommenting \hspace{0.05cm} \emph{``It's difficult to consider all branches of the main principles that should be applied to AI. This results in some gaps and scenarios being missed and only being realised once it is developed."} -- [P63] 

\subsubsection*{\textbf{Cost}}
Cost is an important factor in developing an AI-based system and it could be one of the challenges in developing ethical AI-based systems. The absence of adequate funds could be a challenge because developing such systems requires specialised skills and resources, which come at a cost and it may also impact business goals. Most of the participants (P1, P3, P5, P9, P10, P14, P15, P16, P20, P21, P22, P26, P90) reported that cost is a challenge in developing ethical AI-based systems.\\
\faCommenting \hspace{0.05cm} \emph{``Developing ethical AI requires a specific set of skills and resources, which may be in short supply and requires a cost"} -- [P58]

\subsubsection*{\textbf{{Time}}}
To apply principles: Time is another factor that may impact the development of ethical AI-based systems. Most of the participants noted that lack of time is a challenge they face in considering and following ethics during the development of AI-based systems (P5, P7, P10, P11, P14, P16, P17, P18, P21, P25, P26, P29, P90).\\
\faCommenting \hspace{0.05cm} \emph{``Some ethical principles directly impact the development timeline and require additional work which isn't required for the software to be operational, extending the development time-frame."} -- [P70]

To understand principles: It is essential to understand AI ethical principles before incorporating them into the development of AI-based systems. Some of the participants (P5, P6, P8, P9, P10, P11, P17, P18) reported that lack of time to understand the AI ethical principles is a challenge that impacts the development of ethical AI-based systems.

\subsubsection{Technology-related challenges}
Participants discussed two challenges (concepts) related to technology in incorporating ethics during AI-based system development including (i) AI-related and (ii) Data-related challenges. Each of these concepts is underpinned by multiple codes. 

\subsubsection*{\textbf{AI-related}}
Complexity: AI is a complex technology. AI is designed to learn and adapt to new situations, making it more intelligent over time. However, this also means that AI systems can become incredibly complex, making them difficult to understand and control. The complexity of AI technology could be one of the challenges for AI practitioners to follow ethics during its development. This idea was supported by some of the participants who stated:\\
\faCommenting \hspace{0.05cm} \emph{``Systems based on artificial intelligence pursue a strictly defined goal. Make it an increase in usability. At the outset, it is necessary to describe in a mathematical way what utility is. It's hard to do. It is similar to values‚ such as happiness, satisfaction, and quality of life. How to define them mathematically?"} -- [P72]\\
\faCommenting \hspace{0.05cm} \emph{``The reasoning of the system cannot really be 
accounted for, it acts like a black box, where why it came to a certain conclusion is difficult to discern."} -- [P59]

A few other participants like [P42], [P47], and [P71] also reported the same notion. \\

Scope: AI has the potential to tackle a wide range of complex tasks, from recognising speech and images to analysing large data sets and making predictions based on patterns. The scope of AI has been expanding rapidly in recent years, driven by advances in machine learning, deep learning, and natural language processing which could be a challenge for AI practitioners to develop ethical AI-based systems. A participant [P36] shared a similar thought on AI scope. \\
\faCommenting \hspace{0.05cm} \emph{``AI in itself is not very hard to understand but then the problem lies in the user's knowledge of what the data collection and training results they are experiencing turns into, software misuse also happens but with AI the scope is wider and harder to consider."} -- [P36]

Unpredictability: AI is unpredictable as it is difficult to predict the behavior and outcomes of AI-based systems. AI-based systems are designed to learn from data and adapt to new situations, which can result in unpredictable behavior that is difficult to anticipate or control. This behavior of AI could be a challenge for AI practitioners in developing ethical systems. Some of the participants (P81, P87, P99) shared similar thoughts on the unpredictability of AI-based systems when asked about the challenge of developing ethical AI-based systems. \\
\faCommenting \hspace{0.05cm} \emph{``Often it is not clear even to the designer how the system has generated the outcome, or the biases built in to benefit the product owner are obscured from even some of those tasked with its implementation or use"} -- [P99]\\
 Other participants [P81] and [P87] shared the same idea on AI unpredictability. 

\subsubsection*{\textbf{Data-related}}
Lack of quality data: Quality training data is critical to the success of AI-based systems. In order for AI models to accurately learn and make predictions, they require large amounts of high-quality data. Lack of quality training data could be one of the challenges in developing ethical AI-based systems. For example, if the training data is gender or race-biased, then the AI model created will also be biased. Several participants (P58, P59, P60, P74, P76, P88) reported it as a challenge in developing ethical AI-based systems. \\    
\faCommenting \hspace{0.05cm} \emph{``One of the main challenges in developing ethical AI is ensuring that the data used to train the models is unbiased. If the data is biased, the model will be too, potentially leading to unfair or discriminatory outcomes."} -- [P58] \\
\faCommenting \hspace{0.05cm} \emph{``AI systems can perpetuate and amplify biases and discrimination present in the data used to train them, making it challenging to ensure that the AI is treating all individuals fairly."} -- [P88]

Lack of training data: There should be sufficient training data available to create AI models. Lack of training data could be a challenge in creating ethical AI-based systems. This idea was supported by some of the participants (P65, P87).\\
\faCommenting \hspace{0.05cm} \emph{``In my experience, this would be the lack of data, in turn, resulting in not being fully ethical"} -- [P65]

\subsubsection{Human-related challenges} \label{Findings_Humancentric}
Participants discussed four human-related challenges (concepts)  in considering and following ethical principles during AI-based system development including (i) Lack of common perception, (ii) Lack of knowledge/understanding, (iii) Lack of foresight, and (iv) Nature of humans. Each of these concepts is underpinned by multiple codes. 

\subsubsection*{\textbf{Lack of common perception}}
Lack of common perception of ethics: The concept of `ethics' is subjective, with each person having their own unique interpretation, which can differ from one another. This difference in human perception and varying opinions on what constitutes ethics creates difficulties in creating a universal definition of the term `ethics' that hampers the incorporation of ethics in AI. Some of the participants [P27], [P93], [P71], and [P100] supported this notion.\\ 
\faCommenting \hspace{0.05cm}\emph {``As I have been in conversations where people actually don't agree on ethics in the most basic way."} -- [P71]\\
\faCommenting \hspace{0.05cm} \emph{``Difference in thinking about ethics."} -- [P93]\\
\faCommenting \hspace{0.05cm}\emph{``A major limitation is to do with the problem that implementing ethics in AI is highly variable. Meaning that ``ethics" are dependent on the individual/socially accepted practices and behaviors surrounding humans, not some predefined set of a commonly agreed set of rules or behaviors."} -- [P100] 

Likewise, participants also mentioned that a lack of common perception on selecting ethical principles during AI development is a challenge. For example, P30 said:\\
\faCommenting \hspace{0.05cm} \emph{``The major challenge is to agree upon what is an ethical principle that should be taken into account."} -- [P30] \\
Other participants ([P88] and [P63]) also shared similar thoughts on the challenge of selecting ethical principles during AI development. 

Lack of common perception on how to follow ethics and incorporate them during AI development is also a challenge to AI practitioners. A participant [P34] mentioned:\\ 
\faCommenting \hspace{0.05cm} \emph{``There isn't any standard procedure to follow and everyone has different ideas."} -- [P34]

Cause of lack of common perception of ethics: Individuals come from various backgrounds such as different societies, cultures, ethnicity, and groups. These varied backgrounds and cultural differences can influence one's perception of ethics and shape their values which makes it challenging to define the term `ethics'. This notion was supported by several participants (P27, P37, P46, P72, P75, P79, P80, P97). For example, a participant [P97] stated:\\
\faCommenting \hspace{0.05cm}\emph{``Ethics can vary based on different value systems in humans/cultures, therefore I think a big limitation with implementing ethics in AI would be taking a unified approach and applying it, especially with some ethics in particular, and also the potential for misguided ethics."} -- [P97]

Similarly, [P27] and [P37] discussed a similar idea and stated: \\
\faCommenting \hspace{0.05cm}\emph{``Different societal groups have completely different perceptions in terms of ethical issues, therefore, creating a model which will objectively define these things for the AI will be complex."} -- [P27] \\
\faCommenting \hspace{0.05cm} \emph{``Human values are relative."} -- [P37] \\
\faCommenting \hspace{0.05cm} \emph{``I think human-centered values limit us a lot from reaching our potential. We must always put people forward, but this is the right way to do AI and fairness is also challenging as we deal with different people with different values."} -- [P79]\\

Consequence of lack of common perception of ethics: The definition of the term `ethics' of AI varies between countries and companies. This is caused by the difference in human perception of \emph{ethics}. This variance in definitions can cause confusion for AI practitioners, who may not know which definition to adhere to when developing AI-based systems. The absence of a universal definition for the term `ethics' poses a challenge in applying it to AI-based systems. This notion has been supported by several participants (P27, P42, P45, P49, P58, P60, P88)  who stated:\\ 
\faCommenting \hspace{0.05cm}\emph{``The human limitations involve the general definition and perception as to what ethics actually are."} -- [P27]\\
\faCommenting \hspace{0.05cm}\emph{``Artificial intelligence is a very complex issue, and the lack of a clear definition of ethics makes it difficult to apply in this area."} --[P88]\\
\faCommenting \hspace{0.05cm} \emph{``Coming up with universal guidelines, the ethics depend on the company or the organisation designing and developing the solutions and if they manipulate the idea to be in their favor."} -- [P45]

There is not only a lack of a common definition for the term `ethics' but there is also a variation in the definitions of ethical principles of AI. The list of ethical principles and their definitions can differ between companies and countries. Consequently, AI practitioners find it challenging to determine which ethical principles to adhere to while developing AI-based systems. A participant [P58] supported that lack of consensus on AI ethical principles is one of the human limitations in ethics incorporation in AI and stated:\\
\faCommenting \hspace{0.05cm}\emph{``There is often a lack of consensus on ethical principles, which can make it difficult to develop and implement effective guidelines for AI."} -- [P58]

\subsubsection*{\textbf{Lack of Knowledge/Understanding}}
Lack of knowledge/understanding of AI: Presently, AI technology is advancing at an accelerated pace. However, its complexity can make it challenging for humans to comprehend. Staying up-to-date with the latest AI advancements and updates can also be difficult for them which could be a limitation in incorporating ethics in AI. Some of the participants (P35, P40, P41, P58, P60, P62, P70, P77, P81, P82) noted the same. For example, [P35] and [P82] stated:\\
\faCommenting \hspace{0.05cm} \emph{``Lack of understanding of the technology."} -- [P35]\\
\faCommenting \hspace{0.05cm} \emph{``There are several barriers or constraints to implementing AI, but the main ones for me are the shortage of knowledge and skills and availability of technical personnel with the experience and training necessary to effectively implement and operate ethical AI solutions."} -- [P82] 

Likewise, [P41] also mentioned that humans lack understanding of creating ethical AI systems as AI is powerful and rapidly evolving. \\
\faCommenting \hspace{0.05cm} \emph{``The field of AI ethics is relatively new and rapidly evolving, and there is still much that is not understood about how to create ethical AI systems."} -- [P41]
 
A similar thought was shared by [P58] who said that a lack of full understanding of the ethical implications of AI leads to difficulty in developing and incorporating ethical guidelines. [P58] reported:\\
\faCommenting \hspace{0.05cm} \emph{``Many people may not fully understand the ethical implications of AI, making it difficult to develop ethical guidelines for its development and use."} -- [P58]

A similar idea was discussed by [P70] where the participant said that AI is too advanced and practitioners' lack of knowledge of advanced programming is a limitation to incorporating ethics in AI. The participant stated:\\
\faCommenting \hspace{0.05cm} \emph{``The limitations stem from the advancement in AI to the point where it becomes too advanced for simple programming."} -- [P70]\\

Cause of lack of knowledge/understanding of AI:
AI is a complex and multifaceted technology. Its primary characteristics are self-learning and adaptiveness, which enable it to constantly improve its performance and decision-making abilities. However, its outcomes can be difficult for humans to comprehend, and it can sometimes be seen as a ``black box". The complexity of AI has been a topic of discussion among AI experts, as it is one of the causes of humans' limited understanding of AI. \\
\faCommenting \hspace{0.05cm} \emph{``AI decisions are not always intelligible to humans."} -- [P82]

Many participants mentioned the complexity of AI as one of the main causes of human's lack of understanding of AI (P41, P45, P70, P81). \\

Consequence of lack of knowledge/understanding of AI: AI has gained immense power and can sometimes surpass human capabilities, despite being created by humans. However, due to its intricate nature and humans' limited understanding of AI, there is a potential for ethical vulnerabilities to be exploited by AI. P81 stated: \\
\faCommenting \hspace{0.05cm} \emph{``We can never tell what loopholes are there in our currently existing list of ethics. AI is super intelligent, and there may come a point where it finds a loophole and destroys us. Our limitation is that we do not know everything."} -- [P81]\\

Lack of knowledge of other's work: Nowadays, AI is extensively utilised in various fields like healthcare, transportation, finance, information technology, education, and many others. With the growing prevalence of AI, there is also an upsurge in the creation of AI-based systems. However, since AI practitioners are scattered across the world, they may not always be aware of each other's progress and actions, which can hinder the implementation of ethical practices in AI due to human limitations.   
A participant [P48] shared similar idea:\\
\faCommenting \hspace{0.05cm} \emph{``What limits us, is everyone working on AI independently, not knowing what the next company is doing "} -- [P48]

\subsubsection*{\textbf{Lack of foresight}}
AI outcomes: AI practitioners who developed AI-based systems are unable to predict the outcomes of the system due to the complex nature of AI. The complexity of AI not only causes a lack of understanding of AI in AI practitioners but also makes it challenging to predict AI outcomes for them. Some participants (P36, P52, P59, P88) supported this idea: \\
\faCommenting \hspace{0.05cm} \emph{``Our limits come from our understanding of AI itself, where we do not know all the parameters that it gives out. At the end of the day, we're left to guess the end results of what our AI does and how it will function."} -- [P52]\\
\faCommenting \hspace{0.05cm} \emph{``To consider AI's possibilities outside the scope of human possibilities or the things easily achievable as a human, it is necessary to fully consider the outcome of AI use in any field and this could become a hard task to manage with how AI is progressing."} -- [P36]\\
\faCommenting \hspace{0.05cm} \emph{``Ensuring that AI systems are transparent and accountable for their actions can be difficult, as it can be challenging to understand how an AI system arrived at a particular decision."} -- [P88] \\

AI consequences: Like any other software, AI-based software may have both positive and negative consequences. One of the human limitations is the inability to predict the consequences of AI-based software. Many participants mentioned that lack of foresight of AI consequences is one of the human-related challenges in incorporating ethics in AI (P1, P2, P5, P6, P7, P9, P10, P11, P14, P15, P16, P18, P23, P58, P88, P41, P81, P69). \\
\faCommenting \hspace{0.05cm} \emph{``It can be difficult to predict the long-term consequences of AI, making it challenging to anticipate and mitigate potential negative impacts."} -- [P58]\\
\faCommenting \hspace{0.05cm} \emph {``Difficulty in predicting future consequences"} -[P60]\\
\faCommenting \hspace{0.05cm} \emph{``It can be difficult for developers to anticipate all of the potential consequences of an AI system, particularly if the system is highly complex or if it is being used in a new or unexpected way."} -- [P41] 

\subsubsection*{\textbf{Nature of humans}} 
Biased: Humanity is divided into various categories including culture, ethnicity, country of origin, and religion. This division creates diversity in the values people adopt and how they perceive things, resulting in biases. This is an inherent characteristic of human nature and represents a significant challenge in the development of ethical AI-based systems. The majority of the participants reported it as a major human-related challenge in considering and following ethical principles of AI (P1, P2, P5, P6, P7, P8, P10, P11, P13, P15, P16, P19, P20, P21, P22, P23, P26, P29, P30, P33, P34, P39, P44, P45, P54, P59, P60, P61, P63, P64, P69, P73, P75, P78, P80, P82, P83, P84, P85, P96, P99, P100).   \\
\faCommenting \hspace{0.05cm} \emph {``It's impossible to have a no-bias perspective and be conscious of all the moral implications of our work, even if we're working on a small or mid-size team."} -- [P29]\\
\faCommenting \hspace{0.05cm} \emph{``Humans are inherently biased, and in implementing machine learning systems, some of these biases are found."} -- [P59] \\
\faCommenting \hspace{0.05cm} \emph{``Humans are subject to their own bias which tends to seep into the logic used to build the AI."} -- [P63] \\
\faCommenting \hspace{0.05cm} \emph{``The limitations concern the innate human view of things. A human being, no matter how hard he tries, can never be totally immune to the bias or to the psychological and cultural structures that formed him."} -- [P44]\\
\faCommenting \hspace{0.05cm} \emph{``The main barriers in my opinion would be bias or discrimination, and potentially the philosophical challenge with regard to humans' involvement in developing AI systems."} -- [P80] \\

Awareness of bias transfer: Humans are inherently biased, and there is a significant likelihood that their biases may be transferred to AI-based systems during their development. Even those with good intentions may inadvertently transfer their biases, while those with bad intentions may do so intentionally. Consequently, humans may or may not be conscious of the extent to which their biases have been transferred during the development of AI-based systems. Many participants reported it as a challenge in developing AI-based systems (P38, P44, P63, P90, P94, P96, P99).  \\
\faCommenting \hspace{0.05cm} \emph {``I think the humans (behind the AI) implicit biases may be transferred to the AI be it intentionally or unintentionally."} -- [P38]\\
\faCommenting \hspace{0.05cm} \emph{``The human bias is unknowingly being transferred in the code."} -- [P90] \\
\faCommenting \hspace{0.05cm} \emph{``That our own ethical faults get transferred into the ethics system"} -- [P94] \\

Ethics vs profit: Diverse preferences and priorities are inherent traits of human beings. For instance, certain individuals may prioritise financial gain over ethical considerations, while others may prioritise ethics over profit. This human nature is one of the challenges in developing ethical AI-based systems and similar thought was shared by some of the participants:\\
\faCommenting \hspace{0.05cm} \emph{``Humans, for example, insurance companies, care about profit. Including ethics in an AI, for example, in regard to diversity and inclusion, can be counter-intuitive or counterproductive. If statistics and the AI itself determine that people from a given ethnicity are prone to a given outcome that is adverse to business, it is counter-intuitive for us programmers to not take it into consideration if it affects what we are trying to maximise. Nature is what it is, numbers are what they are, and these things are very difficult to control and maybe, they shouldn't. It is like asking someone to imagine that gravity is non-existent and jump off a roof."} -- [P30]\\
\faCommenting \hspace{0.05cm} \emph{``The danger lies in humans implementing AI for their own gain, and for AI practitioners to ignore ethics for profit, greed, selfishness, or any other negative reason."} -- [P100]\\

\subsection{Summary of Key Findings} \label{Summary_KeyFindings}
The key findings from our survey have been summarised in Table \ref{table:KF}. Our previous work (GTLR) \citep{pant2022AIethics}, revealed the need for an empirical study that solely focuses on investigating AI practitioners' views and understanding of ethics in AI. We, therefore, focused our study on investigating aspects related to AI practitioners'  \emph{awareness} of AI ethics, and their \emph{challenges} in incorporating ethics in AI-based systems.

\begin{table}[ht]
\centering
\caption{Key Findings (KF) of the study.} \label{table:KF}     

\scriptsize
\begin{tabular} {>{\raggedright\arraybackslash}p{0.5cm}>{\raggedright\arraybackslash}p{9.5cm}>{\raggedright\arraybackslash}p{0.5cm}}

\hline\noalign{\smallskip}
 & Key Findings (KF) & Section\\

\hline\noalign{\smallskip}
KF1 & Majority of the AI practitioners (41\%) were \emph{reasonably} familiar with the concept of ethics in AI. & 4.2.1  \\

KF2 & Few AI practitioners (13\%) were very familiar with the concept of \emph{ethics} in AI. &  4.2.1  \\

KF3 &  Majority of the AI practitioners were aware of some AI ethical principles including \emph{`Privacy protection and security'} (18.03\%), \emph{`Reliability and safety'} (14.93\%), and \emph{`Human-centred values'} (14.93\%). & 4.2.2   \\

KF4 &  Very few AI practitioners (3.38\%) were aware of \emph{all} ethical principles of AI. & 4.2.2  \\

KF5 &  The reason for the majority of AI practitioners’ (23.16\%) awareness of AI ethics and ethical principles were \emph{workplace rules and policies}. &  4.2.3   \\

KF6 & A very few AI practitioners (0.74\%) reported \emph{university} as the reason for their awareness of AI ethics and ethical principles. & 4.2.3 \\

KF7 & Majority of AI practitioners (43\%) believed that formal education or training \emph{moderately} help in preparing them to adhere to AI ethical principles during AI development. & 4.2.4 \\

KF8 & Majority of the participants (27\%) reported \emph{`Human-centred values'} as the most challenging ethical principle to adhere to during AI development. & 4.3 \\

KF9 & AI practitioners encountered \emph{General challenges}, \emph{Technology-related challenges}, and \emph{Human-related challenges} in incorporating ethics in AI. & 4.3 \\

KF10 & Majority (41\%) of the AI practitioners reported \emph{biased nature of humans} as a key human-related challenge in incorporating ethics in AI. & 4.3.3 \\

\noalign{\smallskip}\hline
\end{tabular}
\end{table}

\section{Discussion}\label{sec:Discussion}
We now discuss our findings and insights in light of related works.

\subsection{AI Practitioners’ Awareness of Ethics in AI}
\subsubsection{Extent of awareness of `AI Ethics’ concept}
\citet{vakkuri2020just} reported that there is a lack of awareness of AI ethics among AI practitioners. Most studies have focused on understanding if AI practitioners are aware of the concept of AI ethics, and focused on developing tools and methods to raise awareness among AI practitioners \citep{morley2021ethics}. An empirical study by \citet{mcnamara2018does} reported that the ethical guidelines provided by the Association for Computing Machinery (ACM) had minimal influence on software developers, who continued to work in the same way as before and concluded that software practitioners were not well-informed on ethics. Based on that, the Ethically Aligned Design (EAD) guidelines version acknowledged that this could also be true for AI ethics \citep{vakkuri2020just} but there is no research to investigate how familiar AI practitioners are with the concept of AI ethics. This along with the fact that being aware of AI ethics is insufficient; a thorough understanding of the concept is crucial for AI practitioners to ensure that AI development is conducted in a responsible and ethical manner motivated us to explore this topic. As a result, we carried out a survey involving 100 AI practitioners, revealing that most participants (41\%) possess a \emph{reasonable} level of familiarity with the concept of AI ethics. This suggests that there is still a deficiency in the efforts required to enhance awareness of `AI ethics' among AI practitioners.

\subsubsection{Awareness of AI ethical principles}
Various companies, such as Microsoft \citep{Microsoft}, Google \citep{Google}, and IBM \citep{IBM}, have their own ethical guidelines on AI development, outlining the ethical principles that AI-based systems should be developed based on, such as transparency, fairness, privacy, etc. These guidelines serve to steer AI practitioners toward ethical AI development and ensure that the systems they develop align with all these principles. This implies that AI practitioners must be aware of and possess adequate knowledge of these ethical principles of AI before developing them. However, research shows that AI practitioners are aware of only specific ethical principles of AI such as \textit{accountability/responsibility, privacy, fairness}, and \emph{transparency and explainability}. For example, \citet{vakkuri2019ethically} and \citet{mark2019ethics} concluded that AI practitioners in their respective studies were aware of the `transparency' ethical principle of AI. AI developers were aware of the ethical principle of ``fairness" in AI and strived to eliminate any issues related to it \citep{holstein2019improving}. According to \citet{veale2018fairness}, participants understood the importance of accountability in AI systems and took responsibility for any harm caused by their creations. According to \citet{rothenberger2019relevance}, the principle of ``responsibility" was deemed highly relevant and influential among the other ethical principles in AI. \emph{`Privacy'} was another ethical principle that AI practitioners were aware of and extensively discussed, with data and information privacy being a significant concern for organisations \citep{ibanez2022operationalising, ryan2021research}.  

\citet{christodoulou2021democracy} reported that participants discussed the challenges related to \emph{transparency}, \emph{privacy}, \emph{fairness}, and \emph{accountability} only when they were asked about the challenges in addressing ethical issues in AI. It indicates that AI practitioners were either not aware of other ethical principles of AI or they didn't face any challenges related to them.
We reviewed empirical studies that focused on understanding AI practitioners' views on AI ethics in our previous work (GTLR) \citep{pant2022AIethics}. We found that AI practitioners discussed only four ethical principles of AI including \emph{transparency and explainability}, \emph{privacy}, \emph{fairness}, and \emph{accountability/responsibility}.

The studies conducted reveal that AI practitioners discussed only a few ethical principles of AI, including \emph{privacy, accountability, transparency}, and 
\emph{fairness}. However, it is unclear whether the research context and questions asked were responsible for the AI practitioners only discussing these four ethical principles. Our study fills this gap by investigating what AI ethical principles they are aware of.

We used Australia's AI Ethics Principles\footnotemark[\value{footnote}] and our results show that \emph{`Privacy protection and security'} is the ethical principle that most AI practitioners (18.03\%) were aware of. This suggests that, regardless of the specific research context, the majority of AI practitioners possess knowledge about the ethical principle of AI concerning \emph{`Privacy protection and security'}. We also found that only 3.38\% of the participants were aware of \emph{all} ethical principles of AI.   

\subsubsection{Reasons for awareness}
\emph{Organisational pressure} \citep{veale2018fairness}, \emph{laws and regulations} \citep{vakkuri2019ethically}, \emph{personal interest and experience} \citep{ibanez2022operationalising}, \emph{customer complaints} and \emph{negative media coverage} \citep{holstein2019improving} were some of the reasons for AI practitioners' awareness of AI ethics reported in the literature. We consolidated the reasons cited in previous studies and asked survey participants about the reason for their awareness and we included an open-text option at the end to allow participants to share any reasons that were not included in the provided list. \emph{Workplace rules and policies} were cited by the majority of the participants (23.16\%) as the reason for being aware of AI ethics which was discussed in one of the studies \citep{vakkuri2019ethically}. Likewise, most of the participants reported \emph{personal interest and experience}, \emph{news and media}, and \emph{customer complaints} as their reasons for awareness of AI ethics.

On the other hand, previous studies did not mention \emph{first-hand professional experience} as a reason for AI practitioners' awareness of AI ethics, however, our study revealed that it is indeed a significant factor (14.34\%). In addition, according to our findings, a small number of participants (0.74\%) identified \emph{university} as a reason for their awareness of AI ethics, which was not previously reported in the literature.

\subsubsection{Role of formal education/training in preparing AI practitioners to incorporate AI ethics}
It has been highlighted in the literature that the topic of ``AI ethics" must be incorporated into the curriculum to make students aware of the concept of AI ethics \citep{borenstein2021emerging, bogina2021educating, chang2022understanding}. 
Although the importance of formal education/training has been highlighted in the literature, there is a lack of research that shows to what extent formal education or training assists AI practitioners to incorporate ethics in AI. Our survey discovered that the majority of participants (43\%) believe that formal education/training \emph{moderately} aids in preparing them to incorporate ethics in AI. From our results, it can be inferred that formal education or training play a role in preparing AI practitioners to incorporate ethics in AI, but their significance may not be paramount.

\subsection{AI Practitioners’ Challenges in Incorporating Ethics in AI}
Studies reported the challenges of AI practitioners in adhering to specific ethical principles of AI during AI development such as transparency \citep{ibanez2022operationalising,sanderson2023ai}, accountability \citep{sun2019mapping}, and fairness \citep{holstein2019improving}. Through our study, we investigated AI practitioners' degree of challenges in considering and following all ethical principles of AI using Australia's AI Ethics Principles\footnotemark[\value{footnote}]. Our results indicated that majority of the participants (27\%) find \emph{`human-centered values'} the most challenging ethical principle to adhere to. This finding contradicts the findings of other studies as those studies did not report about \emph{`human-centered values'}. In fact, none of the empirical studies discussing the challenges related to specific AI ethical principles mentioned \emph{`human-centered values'} as the most challenging ethical principle to adhere to, rather they discussed challenges related to other ethical principles such as \emph{transparency}, \emph{accountability}, and  \emph{fairness}. However, this discrepancy may be due to the differences in ethical principles across countries or organisations, as we used Australia's AI Ethics Principles\footnotemark[\value{footnote}] in our survey but recruited participants from around the world as discussed in Section \ref{sec:Limitations}.

Likewise, AI practitioners encountered challenges in \emph{conceptualising ethics} \citep{vakkuri2019implementing}, dealing with various tensions and trade-offs between AI ethical principles when AI practitioners had to incorporate specific ethical principles of AI \citep{sanderson2023ai}, translating  AI ethical principles into practice \citep{ibanez2022operationalising}, and addressing issues such as \emph{highly general principles, vague principles, lack of technical understanding} that impacted the development of ethical AI-based systems \citep{khan2022ai}. These findings align with our study, where participants faced similar difficulties in defining and conceptualising ethics due to its subjective and complex nature, incorporating all AI ethical principles into development because there are numerous ethical principles that need to be considered, translating AI ethical principles to practice (see Section 4.3.1) and lacking knowledge of AI systems while developing ethical AI-based systems (see Section 4.3.3).  

Although some of the findings of our study are similar to the previous studies, there are some findings that differ from the previous studies. Specifically, we obtained various challenges that are related to humans (AI practitioners) which impact the development of ethical AI-based systems. For example, \emph{lack of common perception, lack of knowledge and understanding of various aspects (like AI, other's work), lack of foresight, and nature of humans} are the challenges that we explored through our study. 
It is important to understand AI practitioners' limitations because these practitioners are responsible for designing and developing AI systems that have a significant impact on society \citep{orr2020attributions}. \citet{orr2020attributions} proposed that there is a need for further research to investigate the limitations that AI practitioners possess when it comes to incorporating ethics in AI. Since their study was based in Australia, the authors recommended exploring this issue among a broader range of AI practitioners. Therefore, due to the research gap, we were motivated to conduct a survey to determine the overall challenges that AI practitioners face when it comes to developing ethical AI-based systems and \emph{`human-related challenges'} are the new findings that we obtained through our study. 

\subsection{Insights} \label{sec:insights}
From the analysis of the open-ended responses and memos taken while employing the \textit{STGT for data analysis} approach and literature, we have discovered a number of noteworthy \textbf{insights}. These primary findings, along with our observations, may be used as recommendations for future research. 

\subsubsection{Role of a university (based on participants' responses):}
According to the findings of \citet{borenstein2021emerging}, it is important to incorporate the topic of AI ethics in the curriculum to educate students and create awareness about it. Despite the literature emphasising the significance of formal education in universities for educating people about AI ethics, our survey revealed that only a few participants acknowledged the role of universities in making them aware of AI ethics and ethical principles. It seems like very few universities have included the topic of AI ethics in the curriculum to make students aware of the concept.

On the other hand, our study also aimed to investigate the impact of formal education or training, such as university courses, in facilitating adherence to AI ethics and ethical principles. However, our results indicate that very few AI practitioners reported universities as extremely helpful in preparing them in incorporating AI ethics during the development of AI-based systems. Thus, it appears that while universities play an important role in creating awareness about AI ethics, they do not have a significant impact in educating and supporting individuals in adhering to AI ethics and ethical principles during AI development. In our opinion, university courses alone will not be enough to support AI practitioners, and some specialised training or continuous support will be needed to ensure that AI ethics are followed and adhered to. 

\subsubsection{Moral vs ethics (based on participants' responses):}
The term `moral’ is closely related to `ethics’. Essentially, at a basic level, ethics is about what is good or bad and what is right or wrong. \citet{hazard1994law} defined the term `ethics' as \textit{``the norms shared by a group on a basis of mutual and usually reciprocal recognition"}, whereas, the term `moral' was defined as \textit{``the notions of right and wrong that guide each of us individually and subjectively in our daily existence"}. We found that the survey participants used the term \textit{`moral'} frequently when asked about the challenges in incorporating ethics in AI. For example, a participant stated:\\
\faCommenting \hspace{0.05cm} \textit{``Different people may have different opinions on what values/morals should be included in AI, which can make it difficult to implement ethics."} – [P88]\\
Some noted that the absence of a common definition of morality poses challenges to applying ethics in AI. A participant [P46] stated the following:\\
\faCommenting \hspace{0.05cm} \textit{``Your own moral compass can be different from other people, and since the AI can’t feel emotions, that can make the AI seem cold and robotic, and that is something you can’t make the AI feel. We haven’t come as far yet.}” – [P46] 

According to the participants, a person's moral perception has an impact on the incorporation of ethics in AI, demonstrating the interdependence of morality and ethics. \citet{stahl2021ethical} also concluded that the concept of ``ethics" encompasses more complexities beyond that and the same term is employed to encompass diverse aspects of the issue of morality. Further studies can be conducted to dive deep into these terms (\textit{moral} and \textit{ethics}) and how they impact the development of ethical AI-based systems. \\

\subsubsection{Lack of consensus on definitions (based on literature):}
The term \textit{`ethics'} has been defined by different people. For example, \emph{`ethics'} is defined as \emph{“the moral principles that govern the behaviors or activities of a person or a group of people”} \citep{nalini2020hitchhiker}. Likewise, \citet{iacovino2002ethical} defined \emph{ethics} as \emph{``the way an individual behaves and the values they hold."}, whereas \citet{payne2006successful} defined \emph{ethics} as, \emph{``A system of value principles or practices and the ability to determine right from wrong."} There are various definitions available for AI ethics and ethical principles.  \citet{8436265} concluded that despite ongoing academic discourse on the connection between AI and ethics for many years, there is still no widely accepted definition or consensus on what AI ethics entails or how it should be labeled. Despite numerous papers and diverse keywords from various fields regarding AI ethics, it remains a difficult task to define the field accurately. For example, \citet{siau2020artificial} defined AI ethics as, \emph{``the principles of developing AI to interact with other AIs and humans ethically and function ethically in society"}, whereas, \cite{powers2020ethics} defined it as, \emph{``The ability of a machine to behave morally, without invoking its moral motivations."} 

There is also a lack of consensus on the ethical principles of AI. For example, the definition of ethical principles varies in different parts of the world. One of the  ethical principles included in Australia's AI Ethics Principles\footnotemark[\value{footnote}] and European Commission's Ethics Guidelines is related to `Diversity, non-discrimination and fairness'. Australia's AI Ethics Principles\footnotemark[\value{footnote}] defined `Fairness' in AI as, \emph {`AI systems should be inclusive and accessible and should not involve or result in unfair discrimination against individuals, communities, or groups."} On the other hand, the European Commission's Ethics Guidelines\footnote{https://digital-strategy.ec.europa.eu/en/library/ethics-guidelines-trustworthy-ai}  defined `Diversity, non-discrimination, and fairness' as, \emph{``AI systems should consider the whole range of human abilities, skills, and requirements, and ensure accessibility  and should focus on (i) avoidance of unfair bias, (ii) accessibility and universal design and (iii) stakeholder participation"}.

According to our survey findings, the absence of a shared understanding of the term `ethics' and the lack of agreement on the definitions of `AI ethical principles' make it challenging to develop ethical AI-based systems. According to the participants, the main cause of this challenge is the varying perceptions of human beings on those terms. Other causes of the lack of consensus on definitions of 'ethics' and `ethical principles' and their impact on the development of AI-based systems, should be explored in more depth so that mitigation measures may be devised.

\subsection{Recommendations}
Based on the findings from our survey, we offer some recommendations for the AI industry, the AI research community, and AI educators.\\
 
\faDesktop \hspace{0.05cm} {Recommendations for Practice}
\begin{itemize}
    \item {\emph{Emphasising workplace rules and policies}}:
Most participants (23.16\%) perceived that \emph{workplace rules and policies} were the reasons for their awareness of AI ethics and ethical principles as discussed in Section \ref{sec: reasons}. Therefore, we suggest that AI companies create policies that encompass all aspects of AI development, including ethical considerations such as fairness, accountability, transparency, etc. to enhance the awareness of AI ethics among AI practitioners.  AI companies should also make it mandatory for AI practitioners to comply with all company policies to ensure adherence to ethical principles during AI development. 

\item {\emph{Practising diversity and inclusion in the hiring processes}}:
Our survey findings validate that the primary human-related challenge encountered by AI practitioners during ethics incorporation in AI is the ``biased nature of humans" (Section \ref{Findings_Humancentric}). Since the majority of AI practitioners reported it as a challenge, diversity, and inclusion in the hiring panels or  decision-making processes could be a potential solution to mitigate the biased nature of humans to ensure that their teams are representative of the diverse communities that they serve. This may help to reduce the likelihood of biased perspectives being integrated into AI systems.  

\end{itemize}

\faGraduationCap \hspace{0.05cm} {Recommendations for Research}
\begin{itemize}
 \item {\emph{Investigating solutions for the challenges of integrating ethics in AI}}:
A set of challenges that AI practitioners encounter when incorporating ethical principles into AI systems were identified which include \emph{general challenges}, \emph{technology-related challenges}, and \emph{human-related challenges}, as discussed in Section \ref{Findings_RQ2}. AI researchers can concentrate on exploring solutions to overcome these challenges faced by AI practitioners, which will improve the integration of ethics into AI for AI practitioners. In the same vein, we recommend that AI practitioners analyse the challenges they encounter when integrating ethical principles into AI, based on the challenges identified in our study. Doing so will help them gain a better understanding of their strengths and weaknesses when incorporating ethics into AI. 

\item {\emph{Investigate the `human-centered values' principle in-depth}}:
The focus of our survey study was to find out the AI practitioner’s degree of challenges in considering and following specific AI ethical principles during AI development (see Section \ref{sec:General}). Our survey confirmed that \emph{human-centered values} is the ethical principle that is most challenging to consider and follow during AI development among the ethical principles listed in \textit{Australia's AI Ethical Principles}. Future research should focus on investigating the challenges associated with the adherence of \emph{human-centered values} during AI development and design solutions to overcome those challenges. Research could also be conducted to explore the potential trade-offs between different ethical principles and the ways in which human values can be effectively integrated into AI systems.

\item {\emph{Better understanding the human-related challenges that AI practitioners face when incorporating ethics in AI}}:
We identified a set of human-related challenges that AI practitioners face when considering and following ethical principles in AI that include \emph{lack of common perception}, \emph{lack of knowledge/ understanding}, \emph{lack of foresight}, and \emph{nature of humans} (Section \ref{Findings_Humancentric}). Researchers can delve into each of these limitations and explore solutions to help AI practitioners overcome these limitations, ultimately improving the incorporation of ethics in AI. 

\item {\emph{Exploring the impact of human biases on AI development}}:
Future research should concentrate on investigating various aspects related to human biases and their influence on the development of ethical AI-based systems. 

\item {\emph{Exploring reasons for AI practitioners' awareness of specific ethical principles}}:
We found that the majority of the AI practitioners were aware of only a few ethical principles of AI including \emph{`Privacy protection and security'} (18.03\%), \emph{`Reliability and safety'} (14.93\%), and \emph{`Human-centred values'} (14.93\%) and very few practitioners (3.38\%) were aware of \emph{all} ethical principles of AI, as shown in Section \ref{FindingsAwarenessethprinciple}. Researchers can conduct further studies to explore the reasons why AI practitioners are aware of only specific ethical principles of AI and how they may affect the development of ethical AI-based systems.

\end{itemize}

\faBook \hspace{0.05cm} {Recommendations for Education}
\begin{itemize}
\item {\emph{Working toward including the topic of `AI ethics' in a curriculum}}:
Very few participants (0.74\%) mentioned that \emph{university courses} helped them to be aware of AI ethics and ethical principles, as discussed in Section \ref{sec: reasons}. Therefore, we recommend that AI educators should include the topic of ``AI ethics" in their curriculum to ensure that students are aware and knowledgeable about ethical considerations related to AI development. Likewise, effective training programs related to the incorporation of AI ethical principles can be organised to aid students in developing ethical AI-based systems. 

\item {\emph{Enhancing formal education/training programs}}:
In our study, a significant number of participants (43\%) indicated that formal education/training plays a moderate role in equipping AI practitioners with the necessary skills to incorporate ethics into AI practices, as presented in Section \ref{Findings_formaleducation}. We recommend that AI educators concentrate on improving the quality of formal education/training initiatives to effectively assist AI practitioners in incorporating ethics and upholding ethical standards throughout the process of AI development. 
\end{itemize}

\section{Limitations and Threats to Validity} \label{sec:Limitations}
Regarding threats to \textbf{external validity}, though we aimed at getting worldwide participation in our survey by advertising it on social media and the Prolific platform, we couldn't achieve it as the maximum participants were from Africa (28\%) as shown in Table \ref{tab:demo}. Hence, our survey findings may be biased and may not be applicable to the entire worldwide software engineering community. However, such generalisation is not achievable in practice.   
Regarding participants' job roles, we identified 18 different job roles among the participants, ranging from less than 1 year to over 10 years of experience in the AI industry. However, the interpretation of these roles may differ between companies and require detailed explanations. To address this, we provided a set of key responsibilities related to AI development and asked participants to select the activities they are involved in. We also added an open text box for them to add any other roles they are involved in that are not included in the list. The majority of the participants reported that they are involved in data collection (19.1\%) data cleaning (15.6\%) and model training (11.9\%), which indicates that we collected data from our target participants. We did not consider the team size organisation size details, AI software domains, or information on the frequency or amount of performing AI development activities since our main focus was to gain AI practitioners' insights into their understanding of different aspects of ethics in AI, which is a limitation of this study. AI encompasses various fields such as machine learning, natural language processing, data science, etc. We did not ask the participants what type of AI-based software they develop. Thus, further studies can focus on understanding them. To ensure that our survey is manageable for AI practitioners, we added some fundamental demographic questions. Additionally, we decided to investigate more intricate demographic and work-related environments in future research. 

Regarding threats to \textbf{internal validity}, we realised that it is difficult to collect detailed data through a survey, which is a limitation. However, a survey works very well if we aim to collect data from a large number of AI practitioners worldwide. Furthermore, since the target participants were AI practitioners working in companies, we could ask a limited set of questions that can be answered within a reasonable time frame. Therefore, we focused on asking the most important questions to the participants considering our future research studies. For example, we asked an open-ended question only to obtain their insights on the overall challenges or barriers AI practitioners face in incorporating ethics in AI, as it was the main aspect we wanted to dive deep into. We used closed-ended questions for all other areas to get a brief understanding of them. 

Using the STGT method for analysing qualitative data posed some challenges. The first, second, and third authors conducted open coding on the survey's qualitative data, and then compared and discussed their codes. However, there were instances where they disagreed with each other as the coding process can be subjective. Reaching a consensus for some codes proved to be difficult. Nevertheless, in these situations, the authors opted to use the codes that were agreed upon by the majority and incorporated them into the paper. We used this method for analysing our qualitative data from the survey with the aim of obtaining original, relevant, and in-depth findings and valuable insights for future research by using its reflective memo-writing technique. One potential limitation of conducting a survey is the lack of certainty regarding the accuracy of the information provided by the participants. Since our survey was anonymous and we were unaware of the participants' identities, it is uncertain whether the information they provided was truthful. In our survey, a significant number of respondents indicated that they possess 1 to 2 years of experience in the area of AI development. Nonetheless, due to the anonymous nature of the survey, we cannot verify the accuracy of this information as we have no personal information of the participants. We used Australia's AI Ethics Principles to inquire about the awareness of ethical principles among participants from around the world. However, since participants were not limited to Australia, the responses may not be entirely accurate due to varying sets of AI ethical principles across countries, which is a limitation of our study.

Regarding the number of survey participants, although 190 participants started answering our survey questions (as shown in Qualtrics records), only 104 participants completed it. The target participants for the survey were AI practitioners involved in the development of AI-based systems, so we had to exclude 4 participants who were students, teachers, researchers, or who did not have experience in AI development. As a result, we had to include the responses of only 100 participants in our survey. All the authors were involved in all stages except that only the first, second, and third authors were involved in the analysis stage of the qualitative data. For the quantitative data, the first author conducted the analysis and shared it with all the other authors to discuss each followed step and technique. 

After multiple discussions, the team finalised the best methods for presenting the findings of the qualitative study (as described in Section \ref{data analysis}). We had multiple conversations regarding the analysis, findings, and methods for presenting the results in order to minimise bias. There is a potential risk to the research's internal validity when using the payment for the second round of data collection. Nevertheless, we carefully examined previous studies \citep{jiarpakdee2021practitioners} and decided to use Prolific. We included two \textit{attention check questions} in between the survey questions to check if the participants were paying attention while answering the survey. Payments for participants were only approved after confirming that they were part of our target participant group, answered both the \textit{attention check questions} correctly, and provided responses to every question.

\section{Conclusion} \label{sec:conclusion}
Understanding AI practitioners' views on ethics in AI has been highlighted in the literature but the lack of empirical research on investigating AI practitioners' views on AI ethics motivated us to conduct this study. This study contributes to understanding the industry perspective on the \emph{awareness} of ethics in AI and the \emph{challenges} in incorporating ethics into AI-based systems. 

We explored four aspects related to the AI practitioners' \emph{awareness} of ethics in AI through our study. The aspects are (i) the extent to which AI practitioners are aware of the concept of ethics in AI, (ii) the ethical principles of AI that AI practitioners are aware of, (iii) reasons for AI practitioners' awareness of ethics and (iv) AI practitioners' awareness of the role of formal education or training in preparing them in incorporating AI ethics. We captured AI practitioners' insights through closed-ended questions and the data were analysed using descriptive statistics for analysis. Our results show that the majority of the participants are \textbf{moderately} aware of the concept of ethics in AI and \emph{\textbf{privacy protection and security}} is the principle that the majority of the participants are aware of. Our results also indicate that \textbf{workplace rules and policies} play a major role in AI practitioners' awareness of ethics in AI and only a \textbf{few} AI practitioners thought that formal education or training is extremely helpful for them in incorporating ethics in AI. 

Similarly, through an open-ended question, we obtained data on the key challenges that AI practitioners face in incorporating ethics in AI, and through a closed-ended question, we obtained insights on the degree of challenges faced by AI practitioners specific to implementing each AI ethical principle. We analysed the open-text answers using the 
\emph{STGT method for data analysis} and closed-ended answers through descriptive statistics for data analysis and categorised the challenges into three sections which include, \emph{general challenges}, \emph{technology-related challenges}, and \emph{human-related challenges}. We found that the majority of the participants believe that the \textbf{biased nature} of human beings is a major challenge in developing ethical AI-based systems. We also found that the majority of the participants find the incorporation of \emph{\textbf{human-centered values}} the most challenging ethical principle during AI development. This study's results provide valuable insights into the industry's perspective on their \emph{awareness} and \emph{challenges} related to AI ethics and its incorporation. The AI research community will gain a better understanding of how AI practitioners view ethics in AI and the challenges they encounter while considering and following ethical principles during AI development. Additionally, the study identified areas that require further investigation to benefit the industry, and AI practitioners can use these findings to improve their understanding and incorporation of AI ethics during AI development.

\section{Acknowledgments}

Aastha Pant is supported by the Faculty of IT Ph.D. scholarship
from Monash University. C. Tantithamthavorn is partially supported by the Australian Research Council’s Discovery Early Career Researcher Award (DECRA) funding scheme
(DE200100941).

\bibliographystyle{ACM-Reference-Format}
\bibliography{bibfile}

\section{Appendices}
\appendix

\scriptsize
\section{Appendix A: Survey Questions} \label{Appendix A}
\textbf{Section A: Demographic Information}\\
1. What is your current job title?
\begin{itemize}
    \item AI Engineer
    \item AI/ Data Scientist
    \item AI/ML Specialist
\item AI Expert
\item AI/ML Practitioner
\item AI Developer
\item AI Designer
\item Prefer not to answer
\item Others:
\end{itemize}

2. How old are you?
\begin{itemize}
    \item Below 20
\item 20-25
\item 26-30
\item 31-35
\item 36-40
\item 41-45
\item 46-50
\item Above 50
\end{itemize} 

3. How would you describe your gender?
\begin{itemize}
    \item Man
    \item Woman
    \item Prefer to self-describe as:
    \item Prefer not to answer
\end{itemize}

4. What is your country of residence? 

5. What is the highest degree or level of education you have completed?
\begin{itemize}
    \item High School
\item Bachelor’s degree
\item Master’s degree
\item Ph.D. or Higher
\item Prefer not to answer
\item Others:
\end{itemize}

6. What activities are you involved in? Select \textbf{all} that apply.
\begin{itemize}
    \item Model requirements
\item Data collection
\item Data cleaning
\item Data labeling
\item Feature engineering
\item Model training
\item Model evaluation
\item Model deployment
\item Model monitoring
\item Others:
\end{itemize}

7. How many years of experience do you have in AI-based software development?
\begin{itemize}
    \item No experience
\item Less than 1 year
\item Between 1 to 2 years
\item Between 3 to 5 years
\item Between 6 to 10 years
\item Between 11-15 years
\item Between 16-20 years
\item Over 20 years
\end{itemize}

\textbf{Section B: AI Practitioners' Awareness of Ethics in AI}
``Ethics in AI refers to the principles of developing AI to interact with other AIs and humans ethically and function ethically in society.”
\textit{From K. Siau and W. Wang, “Artificial intelligence ethics: Ethics of AI and ethical AI,” Journal of Database Management, vol. 31, no. 2, pp. 74–87, 2020}\\

8. How familiar are you with the concept of ethics as it relates to AI development?

 \begin{itemize}
\item Very familiar
\item Reasonably Familiar
\item Somewhat familiar
\item Not very familiar
\item Not at all familiar
\end{itemize}

9. What made you aware of “ethics in AI”? Select all that apply.
\begin{itemize}
    \item Workplace rules and policies
\item Customer complaints
\item First-hand personal experience (e.g. as a software user)
\item First-hand professional experience (e.g. as an AI practitioner)
\item Through news and media
\item I have a personal interest in this
\item Not applicable
\item Others:
\end{itemize}

\textbf{Attention-check question:} The AI ethics test you are about to take part in is very simple, when asked for the most discussed ethical principle of AI, you must select `Fairness'. This is an attention check. 

Based on the text you read above, which ethical principle have you been asked to enter?
\begin{itemize}
    \item Accountability
    \item Fairness
    \item Contestability
    \item Reliability and safety
\end{itemize}

10. Which of the following AI ethical principles are you aware of? Select \textbf{all} that apply.
\emph{These are a selected list of the majority of the principles considered around the world. (Australia’s AI ethics principles: https://www.industry.gov.au/data-and-publications/australias-artificial-intelligence-ethics-framework/australias-aiethics-principles)}

\begin{itemize}
    \item Accountability: people identifiable and accountable for AI system outcomes
\item Contestability: timely process to allow people to challenge the AI system use/outcomes
\item Fairness: inclusive and accessible system
\item Human-centered values: respect human rights, diversity \& autonomy of individuals
\item Human, societal, and environmental well-being: benefit individuals, society, and environment
\item Privacy protection and security: respect \& uphold privacy rights \& ensure data security
\item Reliability and safety: reliably operate in accordance with their intended purpose
\item Transparency and explainability: transparency \& responsible disclosure to help people understand AI impacts \& engagement
\item All
\item None
\item Others:
\end{itemize}

11. How well do you think your formal education/training prepared you to implement ethics in AI?

\begin{itemize}
    \item Extremely well
    \item Very well
    \item Moderately
    \item Slightly
    \item Not at all
\end{itemize}

\textbf{Section C: AI Practitioners' Challenges of Incorporating Ethics in AI}

12. In your experience, how challenging is it to consider and follow the following ethical principles when developing AI-based software solutions? (Please choose \textbf{one} option for each principle) 
\begin{table}[H]
    \centering
    \scriptsize
    \begin{tabular}{c  c  c c  c c c}
       & Very  & Reasonably  & Somewhat  & Not very  & Not at all  & No experience \\
       Accountability  & o & o & o & o & o & o\\
        Contestability  & o & o & o & o & o & o\\         
          Fairness  & o & o & o & o & o & o\\
           Human-centred values  & o & o & o & o & o & o\\
            Human, societal \& environmental well-being  & o & o & o & o & o & o\\
             Privacy protection \& security  & o & o & o & o & o & o\\
              Reliability \& safety  & o & o & o & o & o & o\\
               Transparency \& explainability  & o & o & o & o & o & o\\
            
    \end{tabular}
    \label{tab:4}
    \caption{Degree of challenges in considering and following AI ethical principles}
    \end{table}

\textbf{Attention-check question:} In Australia's AI ethics principles list, how many ethical principles are included? Please select `8'. This is an attention check.
\begin{itemize}
    \item 6
    \item 7
    \item 8
    \item 9
\end{itemize}

13. In your experience, what are the main challenges or barriers to incorporating ethics in AI? (Open-text answer)

14. Based on your experience, is there anything else about ethics in AI you would like to share? (voluntary)

15. If you would like to participate in an interview on this topic with us, please share your name and email address (voluntary).









\end{document}